\documentclass[10pt,aps,pre,amsmath,amsfonts,amssymb,floatfix,superscriptaddress,notitlepage,nofootinbib,twocolumn]{revtex4-1}
\usepackage[pdftex]{graphicx}

\usepackage{url}
\usepackage{xcolor}

\usepackage{xr}

\newcommand{\beginsupplement}{
        \clearpage
        \setcounter{table}{0}
        \renewcommand{\thetable}{S\arabic{table}}
        \setcounter{figure}{0}
        \renewcommand{\thefigure}{S\arabic{figure}}
     }

\begin{document}

\title{Modeling sequence-space exploration \\ and emergence of epistatic signals in protein evolution}

\author{Matteo Bisardi}
 \thanks{These authors contributed equally to this work.}
\affiliation{Laboratoire de Physique de l'Ecole Normale Sup\'erieure, ENS, Universit\'e PSL, CNRS, Sorbonne Universit\'e, Universit\'e de Paris, F-75005 Paris, France}
\affiliation{Sorbonne Universit\'e, CNRS, Institut de Biologie Paris Seine, Biologie Computationnelle et Quantitative LCQB, F-75005 Paris, France}

\author{Juan Rodriguez-Rivas}
 \thanks{These authors contributed equally to this work.}
\affiliation{Sorbonne Universit\'e, CNRS, Institut de Biologie Paris Seine, Biologie Computationnelle et Quantitative LCQB, F-75005 Paris, France}

\author{Francesco Zamponi}
\thanks{Corresponding authors: E-mails: francesco.zamponi@ens.fr; martin.weigt@sorbonne-universite.fr.}
\affiliation{Laboratoire de Physique de l'Ecole Normale Sup\'erieure, ENS, Universit\'e PSL, CNRS, Sorbonne Universit\'e, Universit\'e de Paris, F-75005 Paris, France}

\author{Martin Weigt}
\thanks{Corresponding authors: E-mails: francesco.zamponi@ens.fr; martin.weigt@sorbonne-universite.fr.}
\affiliation{Sorbonne Universit\'e, CNRS, Institut de Biologie Paris Seine, Biologie Computationnelle et Quantitative LCQB, F-75005 Paris, France}


\begin{abstract}
During their evolution, proteins explore sequence space via an interplay between random mutations and phenotypic selection. Here we build upon recent progress in reconstructing data-driven fitness landscapes for families of homologous proteins, to propose stochastic models of experimental protein evolution. These models predict quantitatively important features of experimentally evolved sequence libraries, like fitness distributions and position-specific mutational spectra. They also allow us to efficiently simulate sequence libraries for a vast array of combinations of experimental parameters like sequence divergence, selection strength and library size. We showcase the potential of the approach in re-analyzing two recent experiments to determine protein structure from signals of epistasis emerging in experimental sequence libraries. To be detectable, these signals require sufficiently large and sufficiently diverged libraries. Our modeling framework offers a quantitative explanation for different outcomes of recently published experiments. Furthermore, we can forecast the outcome of time- and resource-intensive evolution experiments, opening thereby a way to computationally optimize experimental protocols.
\end{abstract}

\maketitle

\section{Introduction\label{sec:Intro}}

In the course of evolution, biological sequences encoding proteins explore functional sequence space. The observable sequence variability between homologous sequences, {\em i.e.}~sequences connected by common ancestry, results from a delicate balance between mutation and selection. Mutations tend to randomize sequences, while natural selection prunes most of those mutations having a deleterious effect on fitness. When analyzing large databases of homologous protein families \cite{mistry2021pfam}, we therefore find sequences with 70-80\% different amino acids, but highly conserved functional and structural properties.

In turn, it is possible to search for statistical patterns in ensembles of homologous proteins \cite{durbin1998biological}, using tools borrowed from statistical inference and unsupervised machine learning, and to relate them to selective constraints acting in these proteins. 
The most prominent signal is conservation; a position in a protein not (or rarely) changing amino acid over extended evolutionary time scales, is likely to play an important role in the protein's function ({\em e.g.}~active sites in enzymes) or for the protein's structural stability ({\em e.g.}~residues buried in the protein core). 

A second type of statistical signal has received a lot of attention during the last decade \cite{de2013emerging,levy2017potts,cocco2018inverse}. The correlations between the amino acids present in pairs of residue positions can be extracted via global statistical models like those used in Direct Coupling Analysis (DCA) \cite{weigt2009identification,morcos2011direct}, Gremlin \cite{balakrishnan2011learning} or PSICOV \cite{jones2012psicov}. This signal of residue-residue coevolution results from epistatic couplings between residues in structural contact in the folded proteins, {\em i.e.}~of residue pairs in direct physical interaction in the three-dimensional structure of the protein, even if possibly located at long distance along the primary amino-acid sequence. Coevolutionary methods, in particular when used as input for structurally supervised deep-learning methods like RaptorX \cite{xu2019distance}, DeepMetaPSICOV \cite{greener2019deep}, AlphaFold \cite{senior2020improved} or trRosetta \cite{yang2020improved}, have recently induced a revolution in protein-structure prediction, reaching unprecedented accuracy in computationally predicted structures close to the accuracy of experimentally determined structures \cite{jumper2021highly}. Hundreds of previously unknown protein structures have been predicted this way \cite{ovchinnikov2017protein,tunyasuvunakool2021highly}.

However, coevolutionary methods rely on the availability of large alignments of homologous but diverged proteins, since they rely on statistical signatures extracted from sequence variability \cite{haldane2019influence}. Recently, two groups have independently asked the question, if experimentally generated sequences can be used instead of natural homologs for contact prediction \cite{fantini2020protein,stiffler2020protein}. To this aim, they have proposed and performed similar experiments. First, starting from a given wildtype sequence, they have iterated several rounds of alternating sequence diversification via error-prone PCR (polymerase chain reaction) \cite{cadwell1992randomization}, and selection for functionality (antibiotic resistance for both experiments). In contrast to traditional directed evolution \cite{arnold1998design,arnold2018directed}, selection was very weak (low antibiotic concentrations), so proteins are not simply optimized for function, but may diversify their sequences while maintaining a certain level of functionality. After a certain number of rounds, the resulting sequence library was sequenced, to provide the data for statistical learning. 

The resulting functional sequence libraries were quite diversified, with typical distances of 10-15\% of the sequence length from the wild-type protein used as a starting point. This is much less than in natural protein families, characterized typically by average distances of 70-80\% between homologs. However, the simultaneous emergence of about 10-40 mutations, and the depth of more than $10^4-10^5$ sequences in the experimentally evolved libraries, could make the detection of epistasis, and thus contact prediction, possible~\cite{fantini2020protein,stiffler2020protein}. 

Interestingly, both teams have run plmDCA \cite{ekeberg2013improved}, or evCouplings based on plmDCA \cite{hopf2019evcouplings}, on the data -- with very different results. While the contact signal in~\cite{fantini2020protein} was quite weak, and mostly concentrated to nearby positions along the sequence, \cite{stiffler2020protein} found a sufficiently accurate contact prediction to enable the subsequent construction of an precise structural model. 

To understand the differences in results given the similarity in approaches, we have developed a modeling scheme, which allows us to simulate protein evolution in a data-driven sequence landscape. Comparison of simulated and experimental data of both experiments shows that our simulations reproduce quantitatively the experimental observations. Furthermore, the simulation scheme allows us to control important parameters of the experiments, like the evolutionary distance from the wildtype in the final evolved library, the sequencing depth of the library, or the strength of selection. We find that our model is able to explain the difference in contact prediction between the two experiments in terms of sequence divergence and sequencing depth.

The agreement between simulations and experiments suggests that our modeling framework allows for a quantitative analysis of important questions about protein evolution, like the mechanism underlying sequence space exploration and the emergence of signatures of epistasis with sequence divergence, cf.~also the related Sequence Evolution with Epistatic Contributions (SEEC) model ~\cite{de2020epistatic}. Beyond such basic questions in evolutionary biology, our framework has also the potential to help in optimizing experimental design. To give an example, our simulations predict that both experiments would have benefited from slightly weaker selection, represented by slightly lower antibiotic concentrations. This would have enabled a faster exploration of the neighborhood of the wildtype sequence and the occurrence of slightly more deleterious mutations, which have a better chance to be coupled by epistasis than the predominantly neutral mutations accepted at strong selection. Such predictions are very interesting, since our computational approach is efficient and can be applied to thousands of protein families, while the experiments are expensive in time and resources. Guiding them to increase the success probability may therefore be an impactful strategy.
For instance, our approach can be used to explore different protocols, such as alternating cycles of strong and weak selection.

\section{Results\label{sec:Results}}

\begin{figure*}
\begin{center}
\includegraphics[width=.85\textwidth]{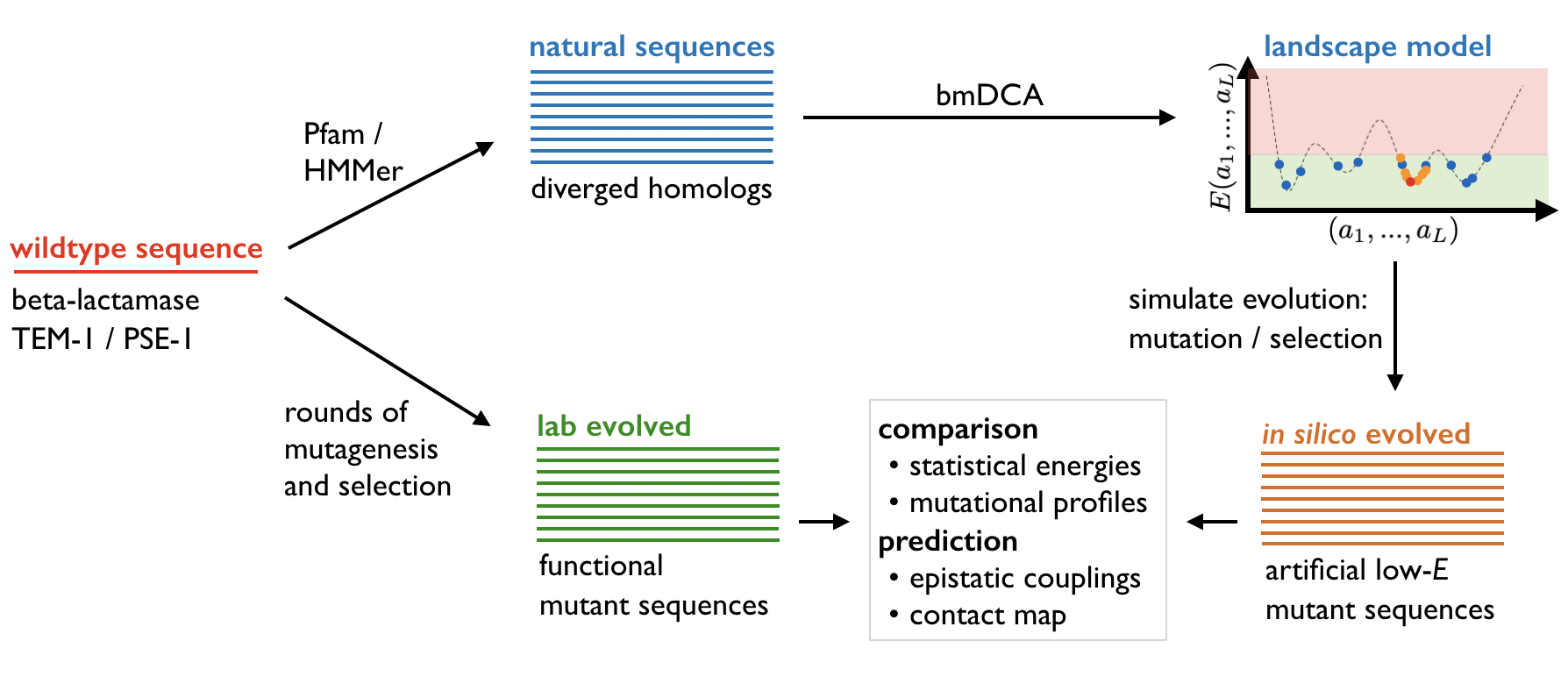}
\vspace*{0.5cm}
\caption{\label{fig:scheme}   {\bf Scheme of our evolutionary modeling approach:} Starting from a wildtype sequence (red), we collect a large multiple-sequence alignment of naturally diverged homologs (blue), which are used to learn a generative landscape model using bmDCA \cite{figliuzzi2018pairwise}. Evolution is simulated as a Markov process in this landscape, leading to simulated, or {\em in silico} evolved mutant sequences. These sequences can be compared to the results of evolution experiments \cite{fantini2020protein,stiffler2020protein} (green), to assess estimated protein fitness (so-called statistical energies, cf. below), mutational profiles, and DCA-based epistasis and contact prediction. The simulation scheme also allows for changing experimental control parameters like final sequence divergence, sequencing depth, and selection strength.}
\end{center}
\end{figure*}

The general procedure of our modeling approach is graphically illustrated in Fig.~\ref{fig:scheme}. In this section, we first describe the data-driven sequence landscape, which is inferred from multiple sequence alignments of natural homologs of the experimentally studied wildtype, {\em i.e.}~from data unrelated to the experiment. As a first check of robustness, we show that this landscape represents well the mutational effects of single-residue substitutions when compared to a deep-mutational scanning experiment, and that the inclusion of epistatic couplings in the landscape model is essential for its accuracy. The landscape can thus be used as a proxy for the protein's fitness landscape.

Next, we present a minimal model of evolutionary dynamics, very similar to but more quantitative than SEEC. In this model, mutations appear at the level of the DNA sequence via single-nucleotide mutations, but selection acts exclusively at the protein level, {\em i.e.} on the amino-acid sequence translated from the DNA sequence, via the inferred sequence landscape. We will show that sequences generated {\em in silico} by this model reproduce quantitative features of the experimentally generated sequences, like mutational profiles or the fitness distribution.

Subsequently, we explore the potential of the experiments by performing simulations under variable conditions for sequence divergence, sequencing depth, or selection strength. This allows us to locate the two experiments in an exhaustively scanned parameter space, to understand the limitations of the experiments, and to propose schemes for overcoming current limitations.

\subsection{An epistatic data-driven sequence landscape captures mutational effects}

The basis of our approach is a computationally inferred sequence landscape, used as a proxy to quantify protein fitness and selection acting on proteins. To obtain this landscape, we first use the Pfam protein-family database \cite{mistry2021pfam} to extract a multiple sequence alignment (MSA) of diverged homologs of the wildtype protein used in the experiments. Both studies performed experiments with a member of the beta-lactamase family (Pfam accession PF13354), TEM-1 in \cite{fantini2020protein} and PSE-1 in \cite{stiffler2020protein}; the latter work also studied the acetyltransferase AAC6 (PF00583). The details of the MSA construction are given in {\em Methods} below; we find, {\em e.g.}, an MSA of 18,334 beta-lactamase sequences.

The underlying idea of our work is to represent the natural variability of this MSA via a generative statistical model $P(a_1,...,a_L)$, with $(a_1,...,a_L)$ representing an aligned amino-acid sequence, {\em i.e.}~the $a_i$ are either one of the 20 natural amino acids, or an alignment gap. Since data are limited, we need to assume some mathematical form for $P(a_1,...,a_L)$. Introducing 
\begin{equation}
    P(a_1,...,a_L) = \frac 1Z \exp\left\{ -E(a_1,...,a_L)
    \right\}\ ,
    \label{eq:P}
\end{equation}
we write the "statistical energy" $E(a_1,...,a_L)$, which is to be seen as a proxy for negative protein fitness \cite{levy2017potts,morcos2014coevolutionary}, in the form used by DCA \cite{weigt2009identification,morcos2011direct,cocco2018inverse},
\begin{equation}
    E(a_1,...,a_L) = - \sum_i h_i(a_i) - \sum_{i<j} J_{ij}(a_i,a_j) \ ,
    \label{eq:E}
\end{equation}
as a sum over position- and amino-acid specific single-residue biases, or fields, $h_i(a_i)$ and pairwise epistatic residue-residue couplings $J_{ij}(a_i,a_j)$. This model, also known as Potts model, assigns low statistical energy $E$ to "good/fit" sequences of high probability, and high $E$ to "bad/unfit" non-functional sequences of low probability. As illustrated in Fig.~\ref{fig:scheme}, we expect to find low statistical energies for both natural and experimentally evolved sequences. The strongest couplings are known to be related to residue-residue contacts in the three-dimensional protein structure, cf.~\cite{morcos2011direct}.

The model parameters are inferred by the currently most accurate version of DCA, called bmDCA \cite{figliuzzi2018pairwise}, which maximizes the model's likelihood via Boltzmann-machine learning \cite{ackley1985learning}. As is known from the literature \cite{sutto2015residue,figliuzzi2018pairwise,levy2017potts}, this model is generative because sequences sampled from $P(a_1,...,a_L)$ reproduce many statistical properties of the MSA of natural sequences.
This does not only concern fitted quantities like one- and two-site amino-acid frequencies, but also non-fitted properties like three-residue frequencies or the clustering of beta-lactamases into subfamilies in sequence space. Note that the epistatic couplings are essential for the model to be generative: a profile model having only fields $h_i(a_i)$ but no couplings $J_{ij}(a_i,a_j)$, {\em i.e.}~a model assuming statistical independence of all positions in the protein, is not generative in the rather strict sense discussed above \cite{figliuzzi2018pairwise}. It misses both non-trivial second- and higher-order correlations and the clustered sequence distribution. Note also that, in a different protein family (chorismate mutase, PF01817), the same modeling approach was recently shown to artificially generate fully {\em in vivo} functional protein sequences \cite{russ2020evolution}.

\begin{figure*}
\begin{center}
\includegraphics[width=\textwidth]{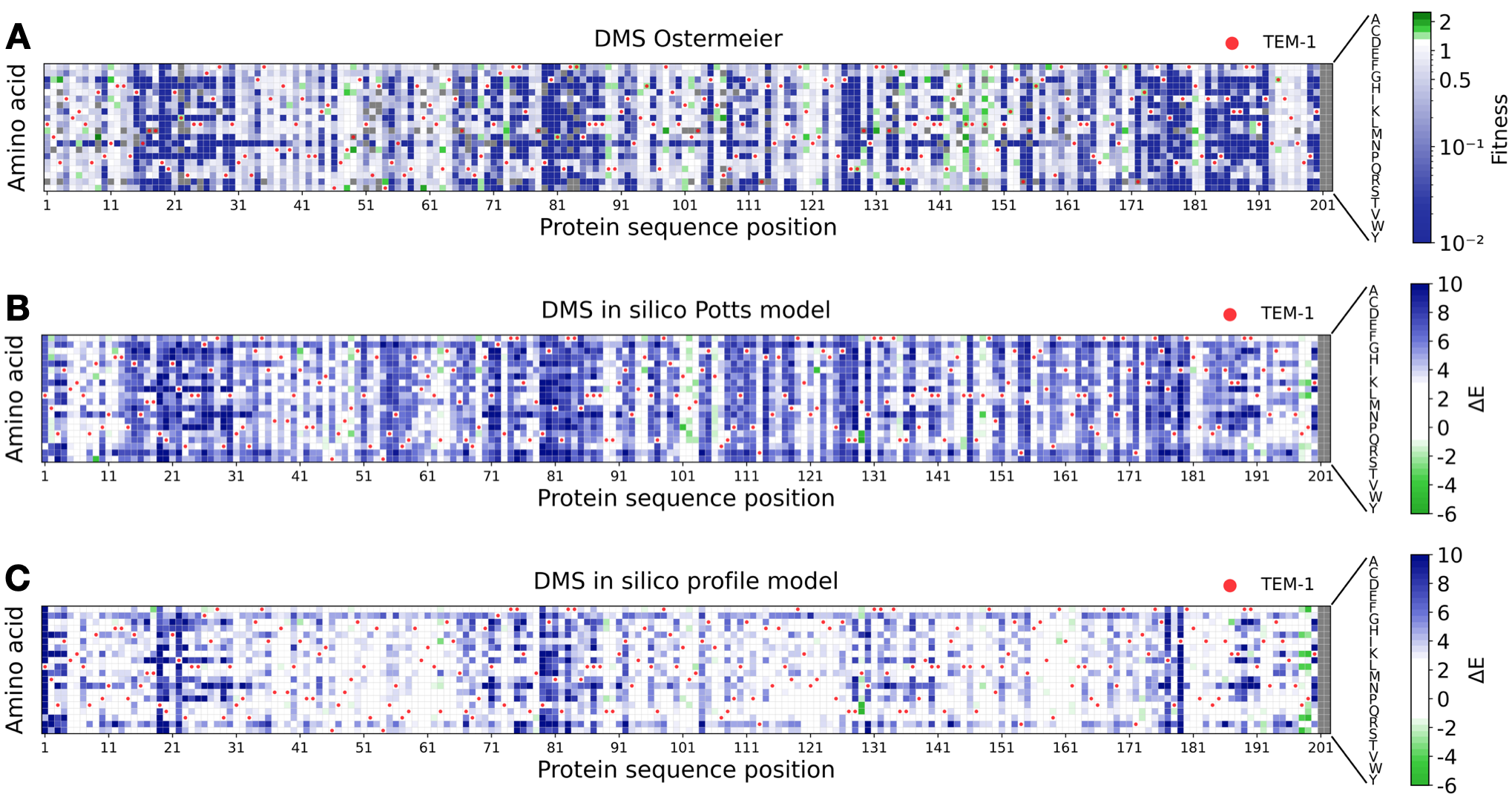}\vspace*{0.5cm}
\caption{\label{fig:dms}   {\bf Experimental and predicted mutational effects in TEM-1:} Panel A shows the results of the deep-mutational scanning experiment of Firnberg et al. \cite{firnberg2014comprehensive}, as compared to the computational predictions using the epistatic Potts model (Panel B) and the non-epistatic profile model (Panel C). Panels A and B have a Spearman rank correlation of -0.77, showing that low energies correspond to high fitness. Panels A and C have a reduced Spearman correlation of -0.6 due to the absence of epistatic couplings in the profile model.}
\end{center}
\end{figure*} 

\begin{figure*}[htb!]
\begin{center}
\includegraphics[width=0.8\textwidth]{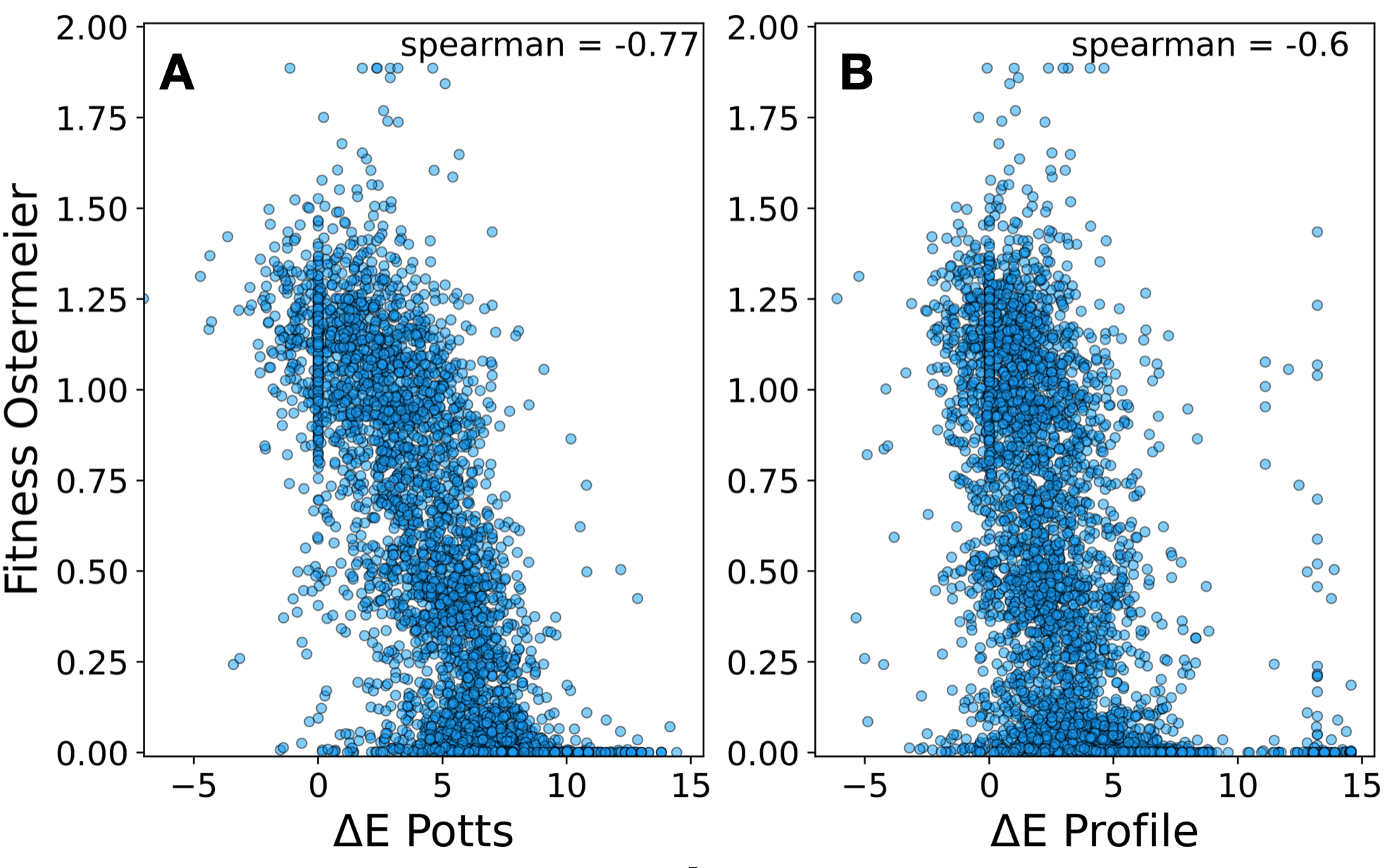}\vspace*{0.5cm}
\caption{\label{fig:dms_scatter} {\bf Predicted vs. experimentally measured mutational effects for TEM-1:} Scatter plot of the data in Fig.~\ref{fig:dms}. Panel A shows the experimental results of Ostermeier et al. vs. the DCA predictions using the epistatic Potts model, Panel B vs. the non-epistatic profile model. The Spearman rank correlations between experiments and predictions are displayed in the figures.}
\end{center}
\end{figure*}

To test the quantitative character of our landscape $E$, we compare the model predictions $\Delta E = E({\rm mutant}) - E({\rm wildtype})$ for the effect of mutations introduced into a wildtype sequence, with the results of a deep-mutational scan of the beta-lactamase TEM-1 \cite{firnberg2014comprehensive}. As is shown in Fig.~\ref{fig:dms}A-B, the two are highly correlated, with a Spearman rank correlation of -0.77, cf.~also~\cite{figliuzzi2016coevolutionary,hopf2017mutation} and the scatter plot supplementary Fig.~\ref{fig:dms_scatter}A directly comparing prediction and experiment. This correlation shows that our landscape $E(a_1,...,a_L)$, even if inferred using distantly diverged TEM-1 homologs, provides quantitative information in the direct vicinity of TEM-1. As expected, low statistical energies correspond to high fitness values. To underline the importance of the epistatic couplings in our model, we also show in Fig.~\ref{fig:dms}C and supplementary Fig.~\ref{fig:dms_scatter}B the predictions of a non-epistatic profile model inferred from the same beta-lactamase MSA: the correlation with the experimental data decreases to -0.6, cf.~\cite{figliuzzi2016coevolutionary}. 

This observation is central for our evolutionary model since the selection of sequences with few mutations with respect to the wildtype reference will be modeled by energy differences $\Delta E$ as introduced above.

\subsection{A model of evolutionary dynamics reproduces quantitative features of experimentally evolved sequences}

Evolution (natural and experimental) can be seen as a stochastic process in a sequence landscape, with random mutations and phenotypic selection modeled by our statistical energy $E(a_1,...,a_L)$. A minimal model realizing this idea is SEEC \cite{de2020epistatic}: a random site $i\in\{1,...,L\}$ is selected, and an amino acid $b\in\{A,C,...,Y\}$ is selected to substitute $a_i$ with a probability proportional to $\exp\{ -\Delta E(a_i\to b) \}$, with $\Delta E$ being the statistical-energy difference between the mutated and the unmutated sequences. A non-accepted or synonymous mutation is characterized by $a_i=b$. Note that deletions and insertions are currently not considered in our model. 

While this model can be used to explore the qualitative influence of epistasis on protein sequence evolution, our analysis requires a more quantitative model taking in particular two differences into account:
\begin{itemize}
    \item Mutations happen at the {\em nucleotide} level. As a consequence, not all amino acids are accessible from all amino acids via a single nucleotide mutation; and the set of accessible amino acids depends specifically on the used codon.
    \item The experiments allow to {\em vary selection strength}. For TEM-1 and PSE-1 this is done by modifying the antibiotic concentration: the same mutation can be more or less strongly favored or suppressed. 
\end{itemize}
To include these factors into our evolutionary model, we introduce two important modifications with respect to SEEC: First, we model evolution at the level of the nucleotide sequence $(n_{11},n_{12},n_{13},...,n_{i1},n_{i2},n_{i3},...,n_{L1},n_{L2},n_{L3})$ coding for the amino-acid sequence $(a_1,...,a_L)$, {\em i.e.} the nucleotide triplet $(n_{i1},n_{i2},n_{i3})$ codes for amino acid $a_i$. For each possible codon $(n_1,n_2,n_3)\in \{A,C,G,T\}^3$ (with the exception of the stop codons), we introduce the set of amino acids ${\cal A}_{acc}(n_1,n_2,n_3) \subset \{A,...,Y\}$, which are accessible from $(n_1,n_2,n_3)$ by a single nucleotide mutation. Possible substitutions for $a_i$ are now only selected from ${\cal A}_{acc}(n_{i1},n_{i2},n_{i3})$, and associated to a single nucleotide change. Note that also $a_i$ is in  ${\cal A}_{acc}(n_{i1},n_{i2},n_{i3})$, accessible via any synonymous mutation.

Second, selection strength will be regulated by a new parameter $\beta$, having the form of an inverse temperature $\beta=1/T$ in statistical physics, which modifies the sequence probability to $P\sim \exp\{-\beta E\}$. The "low-temperature" case $\beta>1$ (${T<1}$) corresponds to increased selection ({\em e.g.} higher antibiotic concentration, or directed evolution), in the limit $\beta\to\infty$ (${T\to 0}$) only the best possible amino acid in position $i$ is accepted. The "high-temperature" case ${\beta<1}$ (${T>1}$) corresponds to decreased selection ({\em e.g.} lower antibiotic concentration); the limit ${\beta\to 0}$ (${T\to\infty}$) describes the case of mutation-accumulation experiments without selection.

This idea is implemented in the following three steps, which are performed recursively, cf.~{\em Methods} for details: 
\begin{enumerate}
\item We randomly select a site $i \in \{1,...,L\}$ to be mutated, corresponding to the codon ${\bf n}_i=(n_{i1},n_{i2},n_{i3})$ and the amino acid $a_i$.
\item One of the accessible amino acids ${b\in {\cal A}_{acc}({\bf n}_i)}$ is selected to substitute $a_i$ with a probability $P(b|a_1,...,a_{i-1},a_{i+1},...,a_L) \propto \exp\{ -\beta \Delta E(a_i\to b) \}$. Due to the epistatic couplings in Eq.~\eqref{eq:E}, this probability depends explicitly on the sequence context $(a_1,...,a_{i-1},a_{i+1},...,a_L)$.
\item One out of the possible codons for amino acid $b$, which differs from ${\bf n}_i$ in a single nucleotide, is selected uniformly at random.
\end{enumerate}
The resulting nucleotide and amino-acid sequences remain thus mutually consistent.

\begin{figure*}[htb!]
\begin{center}
\includegraphics[width=0.8\textwidth]{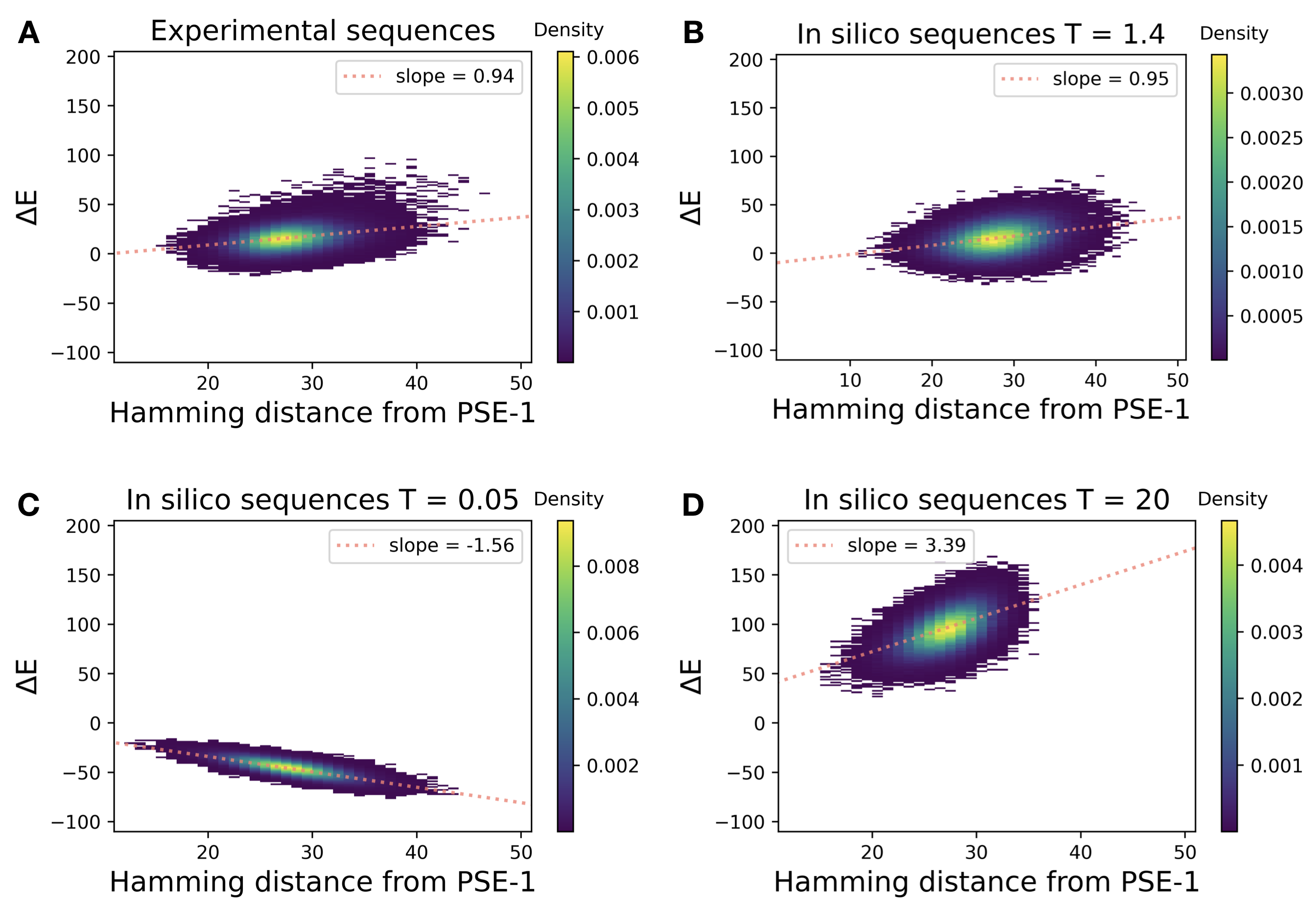}\vspace*{0.5cm}
\caption{\label{fig:E_of_d}   {\bf Statistical energy in dependence of sequence distance from wildtype:} Panel A shows the statistical energies of the sequences from generation 20 in Stiffler et al., as a function of the Hamming distance (number of substituted amino acids) from the wildtype PSE-1. Panel B shows the same quantities for the {\em in silico} simulated sequences, where selection strength $T$ and the number of simulated evolutionary steps are adjusted to reproduce the average distance and the slope from Panel A. Panel C shows an example of strong selection ($T\ll 1$) leading to optimized sequences having lower statistical energies / higher fitness. Panel D shows the case of very weak selection ($T\gg 1$) resulting in random, mostly deleterious substitutions strongly increasing statistical energy.}
\end{center}
\end{figure*} 

The proposed dynamics can be efficiently implemented, and very large sequence libraries can be simulated over long times. To make these data comparable to the libraries generated by experimental evolution, we need to adapt the simulation parameters: first, the number of mutational steps in our simulation is not directly related to the number of experimental generations (because error-prone PCR may introduce multiple mutations each round); we choose it to reach the same average number of substituted amino acids in the simulated and experimental libraries. In this sense, different experimental mutation rates can be parametrized by the number of steps needed by our dynamics to reach the same number of mutations. Second, the selection strength ${\beta=1/T}$ has no evident relation to the antibiotic concentration used in the experiment. We therefore tune the value of $\beta=1/T$ such that the statistical energy $E(a_1,...,a_L)$ of the simulated and the experimental sequences have the same linear slope as a function of the number of substitutions. For the case of PSE-1, shown in Fig.~\ref{fig:E_of_d}, we find that $T=1.4$ is a good value, cf.~Panel A for the experimental data from \cite{stiffler2020protein}, and Panel B for simulated data. This corresponds to low selection strength $\beta=1/T<1$. Even if we adjust only average distance and slope, we find that also the overall distribution is well reproduced. Similar observations for TEM-1 and AAC6 are shown in supplementary Figs.~\ref{fig:E_of_d_tem1} and \ref{fig:E_of_d_aac}.

Fig.~\ref{fig:E_of_d}C shows that for strong selection ${T=0.05}$ (${\beta = 20}$) the sequence energy decreases with the number of substitutions, corresponding to an increasing fitness as expected in a directed-evolution scenario. Weak selection, shown in Fig.~\ref{fig:E_of_d}D for $T=20$ ($\beta = 0.05$), corresponds to a sharp increase in statistical energy, and thus a loss in fitness, as expected from the accumulation of predominantly deleterious random mutations.

\begin{figure*}[htb!]
\begin{center}
\includegraphics[width=\textwidth]{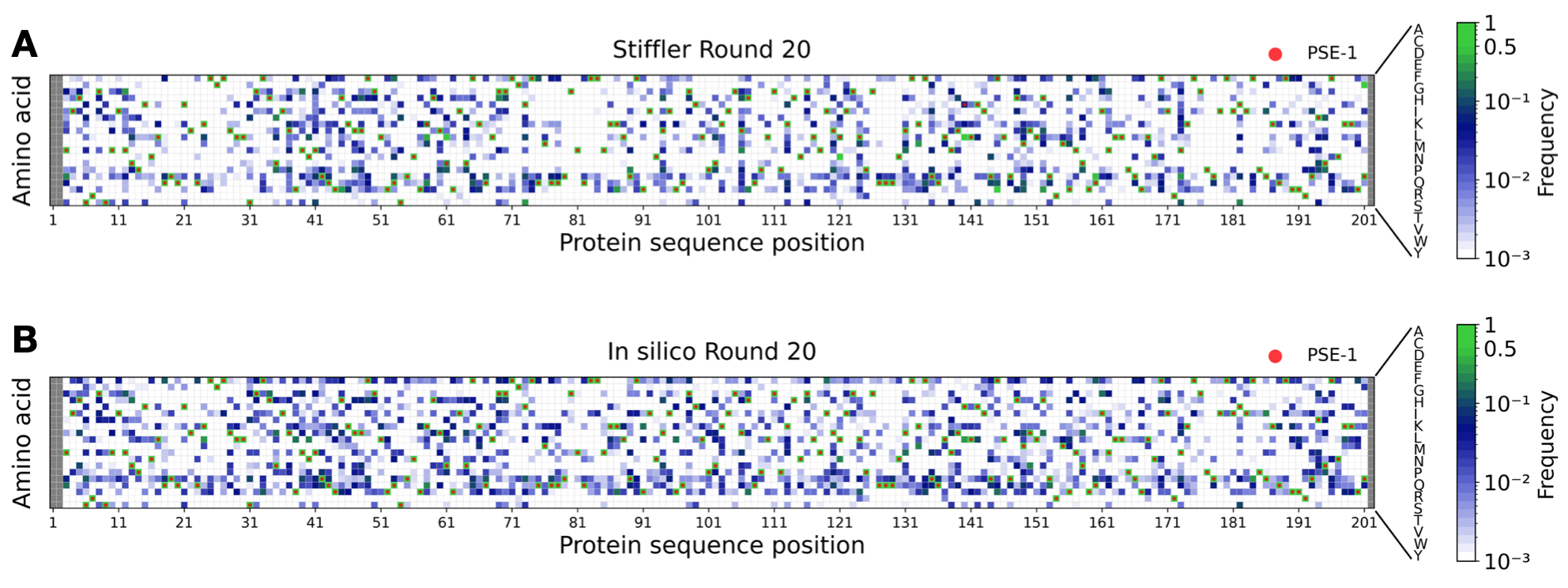}\vspace*{0.5cm}
\caption{\label{fig:mutation_spectrum}   {\bf Position-specific amino-acid frequencies for experimental and simulated sequence libraries:} Panel A shows the frequencies $f_i(a)$ of usage of amino acid $a$ in site $i$ in round 20 of experimental PSE-1 evolution, Panel B shows the same quantity for simulated evolution. 
The Spearman rank correlation between the two frequency spectra is 86\%.}
\end{center}
\end{figure*} 

\begin{figure*}[htb!]
\begin{center}
\includegraphics[width=0.8\textwidth]{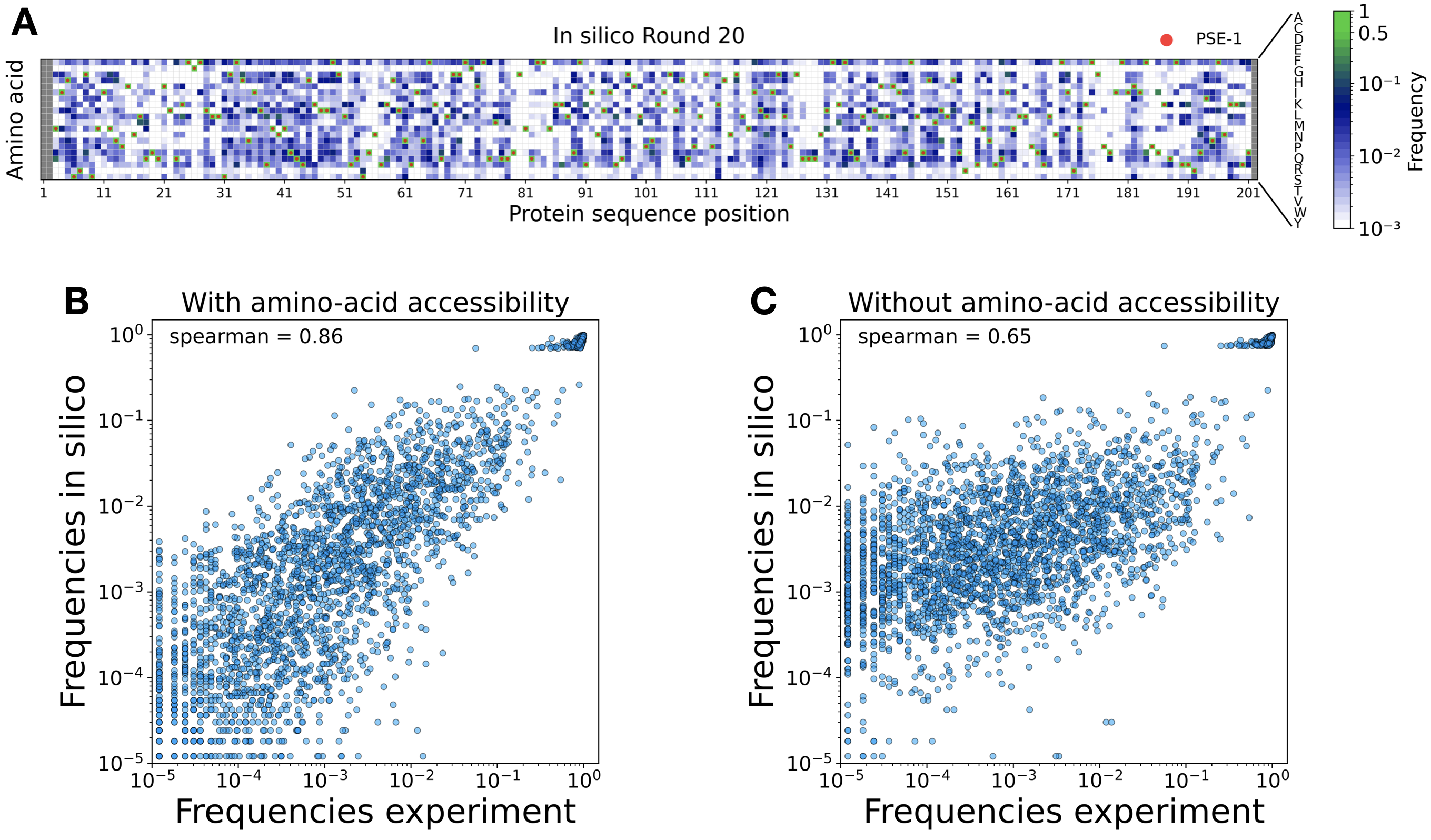}\vspace*{0.5cm}
\caption{\label{fig:mutation_spectrum_S}  {\bf Position specific amino-acid frequencies for experimental and simulated sequence libraries for PSE-1:} Panel A shows the frequencies $f_i(a)$ of usage of amino acid $a$ in site $i$ for the simulated sequences without taking into account amino-acid accessibility. Panel B and C show scatter plots of these frequencies for the experimental data vs. simulated data. Panel B takes amino-acid accessibility into account, and shows a higher correlation than Panel C not taking accessibility into account.}
\end{center}
\end{figure*}

Fig.~\ref{fig:E_of_d}A-B shows global measures comparing experimental and simulated sequences: the Hamming distance is the number of substitutions along the entire amino-acid sequence, the energy also depends on the entire sequence. To increase our confidence in the quantitative character of our evolutionary model, we compare in Fig.~\ref{fig:mutation_spectrum} the site- and amino-acid specific mutational frequencies between experimental and simulated sequence data. To this end, we extract the quantities $f_i(a)$ describing the fraction of sequences in an MSA having amino acid $a$ in position $i$. Interestingly, also this refined measure of sequence diversity is very similar for simulated and experimental sequences; we observe a high correlation of 86\%, cf.~Fig.~\ref{fig:mutation_spectrum_S}. These plots highlight the importance of working only with amino acid substitutions accessible via single-nucleotide mutations: many amino acids show zero frequency in both plots due to inaccessibility. The mutational spectrum predicted without considering the accessibility of amino acids is shown in supplementary Fig.~\ref{fig:mutation_spectrum_S}: we see that the mutational frequencies are more homogeneously distributed, close-to-zero-frequency mutations become very rare as compared to the experimental sequences. The correlation goes down to 65\% between simulated and experimental data in this case.

Based on these observations, we conclude that our evolutionary model, which combines mutations at the nucleotide level with selection at the amino-acid level, is able to reproduce well the statistical features of the experimental sequences. This conclusion is also confirmed, when using TEM-1 and AAC6 as initial wildtype sequences, cf.~Figs.~\ref{fig:mutation_spectrum_TEM1} and \ref{fig:mutation_spectrum_AAC}.


\subsection{{\em In-silico} sequence-space exploration, and the emergence of epistatic signals}

Having developed a quantitative model to simulate experimental evolution, we are now able to explore evolutionary scenarios going well beyond those realized in the experiments. We can systematically analyze the influence of the sequence divergence from wildtype, of the sequenced library depth, and of the selection strength on the accuracy of coevolution-based contact prediction. Each setting of these parameters would require long experiments and would sometimes be inaccessible due to the high number of experimental rounds or the depth of the sequenced library.

\begin{figure*}[htb!]
\begin{center}
\includegraphics[width=\textwidth]{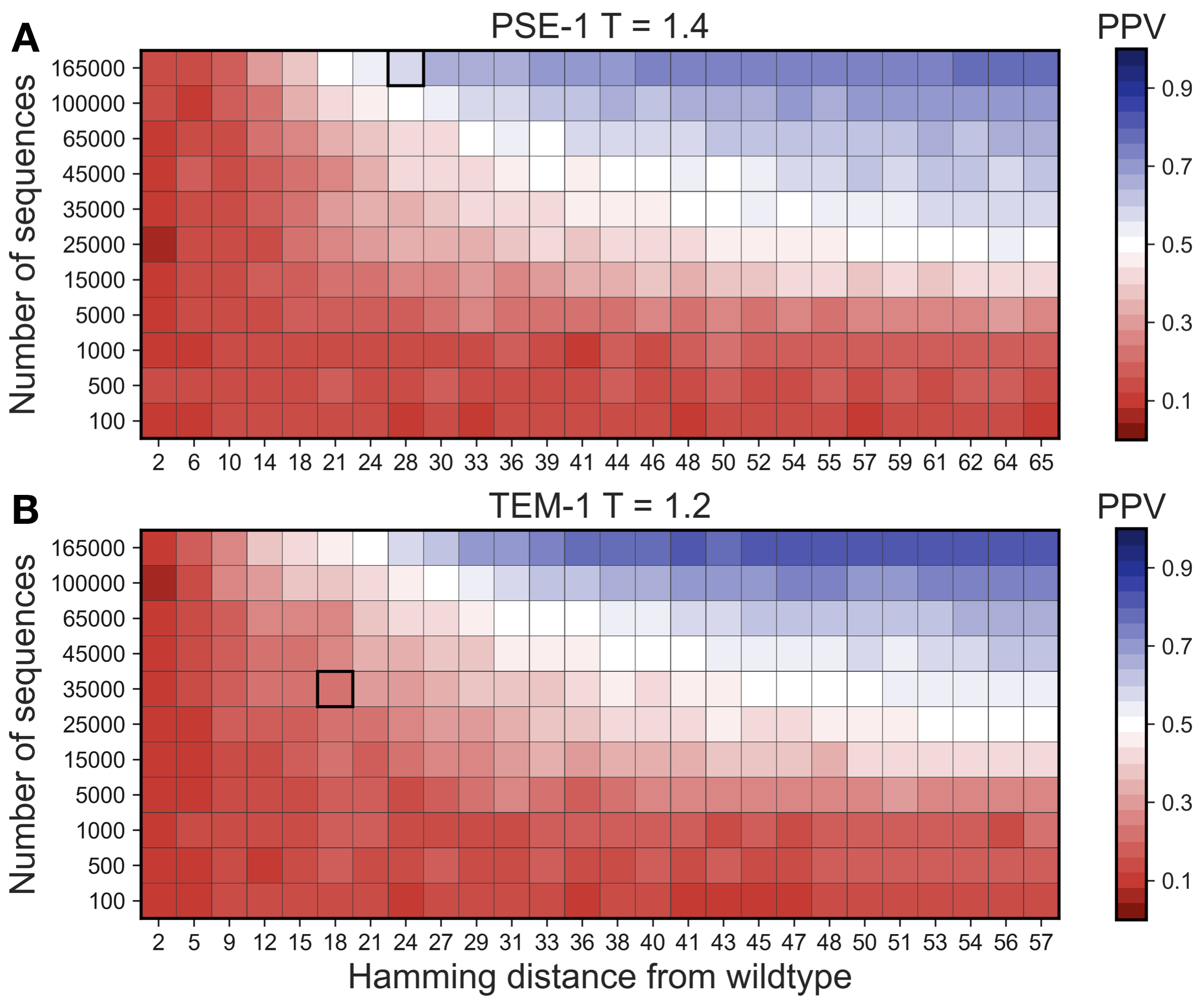}\vspace*{0.5cm}
\caption{\label{fig:phase_diag}   {\bf Accuracy of contact prediction as a function of sequence number and sequence divergence:} Panel A shows the accuracy of contact prediction as a function of the average sequence divergence from wildtype PSE-1 and the depth of the sequenced library. The accuracy is measured via the positive predictive value (PPV), {\em i.e.}, the fraction of true positive contact predictions in the first $100$ DCA-predicted contacts, cf.~{\em Methods} for details. The selection strength $T=1.4$ corresponds to the experimental condition in~\cite{stiffler2020protein}. The highlighted square indicates an average Hamming distance of about 27 and a sequence library of 165,000, as realized in \cite{stiffler2020protein}. Panel B shows the same quantities for wildtype TEM-1, and for the experimental conditions used in \cite{fantini2020protein}. }
\end{center}
\end{figure*} 

Computationally this becomes straightforward although intensive: we have performed many runs of evolutionary simulations, each producing an MSA with specific parameters, simulating the possible outcome of an evolutionary experiment, as represented in Fig.~\ref{fig:phase_diag}. Each square in these plots corresponds to the average over five simulation runs. Depicted is the positive predictive value, which measures the fraction of true positive contact predictions within the first 100 contact predictions, cf.~{\em Methods} for details. Due to the large number of contact predictions to be performed, we used GaussDCA \cite{baldassi2014fast}, a very fast, even if not the most accurate contact predictor. Panel A shows the plot for the selection strength used in the experiments for PSE-1. The red zone corresponds to inaccurate contact predictions, being sometimes hardly better than random (PPV~$\sim$~0.13). It is found consistently for small sequence libraries, and for sequence libraries of low divergence from wildtype. It becomes evident that we need to go to a sufficient number of simultaneous mutations to be able to detect at least a weak epistatic signal between mutations, which can be used for contact prediction. However, this signal remains weak: we need much larger sequence libraries of at least about 50,000 sequences to reach a reasonable contact prediction. However, even for the largest and most diverged library we have studied, a PPV of only 0.7-0.8 is reached, which remains below the contact prediction reached by using the MSA of natural homologs, which was used before for the inference of our sequence landscape. The latter reaches a PPV of $0.98$ using GaussDCA. Panel B shows the same observables for experiments starting with the TEM-1 sequence, the overall results are very similar to PSE-1, even if some quantitative details depend on the initial wildtype sequence. 

It might be speculated that better contact-prediction algorithms may shift the region of non-trivial predictions down to lower Hamming distances from wildtype, or to lower sequence numbers. While the computational cost of plmDCA is too high to reproduce the full analysis of Fig.~\ref{fig:phase_diag}, we have re-analyzed two columns at average Hamming distance $41$ and $65$. As is shown in Supplementary Fig.~\ref{fig:phase_diag_plm}, for low sequence numbers GaussDCA and plmDCA give very similar low prediction accuracies, while the improved accuracy of plmDCA over GaussDCA becomes visible only at sufficiently high sequence numbers. At the resolution of our analysis, no shift in the boundary is observable.

The conditions of the experiments for PSE-1 and TEM-1 are highlighted, in the two panels of Fig.~\ref{fig:phase_diag}. For PSE-1, 20 rounds of evolution led to an average sequence distance of 27 amino-acid substitutions from wildtype, and a sequenced library of 165,000 distinct sequences \cite{stiffler2020protein}. Interestingly, this point is located slightly beyond the boundary of emergence of coevolutionary signal. The predicted average PPV of $0.58$ is comparable to the $0.65$ obtained using the experimental MSA cf.~{\em Methods}.

This is in contrast to the TEM-1 experiment of \cite{fantini2020protein}, cf.~Fig.~\ref{fig:phase_diag}B: the experiment was performed for fewer rounds, leading to less divergence from TEM-1, and the sequence library was less deeply sequenced. The resulting library, with an average Hamming distance of 18 from TEM-1 and with 34431 unique sequences, is located slightly below the line of emergence of coevolution signal. This observation provides a potential explanation for the observed reduced performance in contact prediction.

\begin{figure*}[htb!]
\begin{center}
\includegraphics[width=0.8\textwidth]{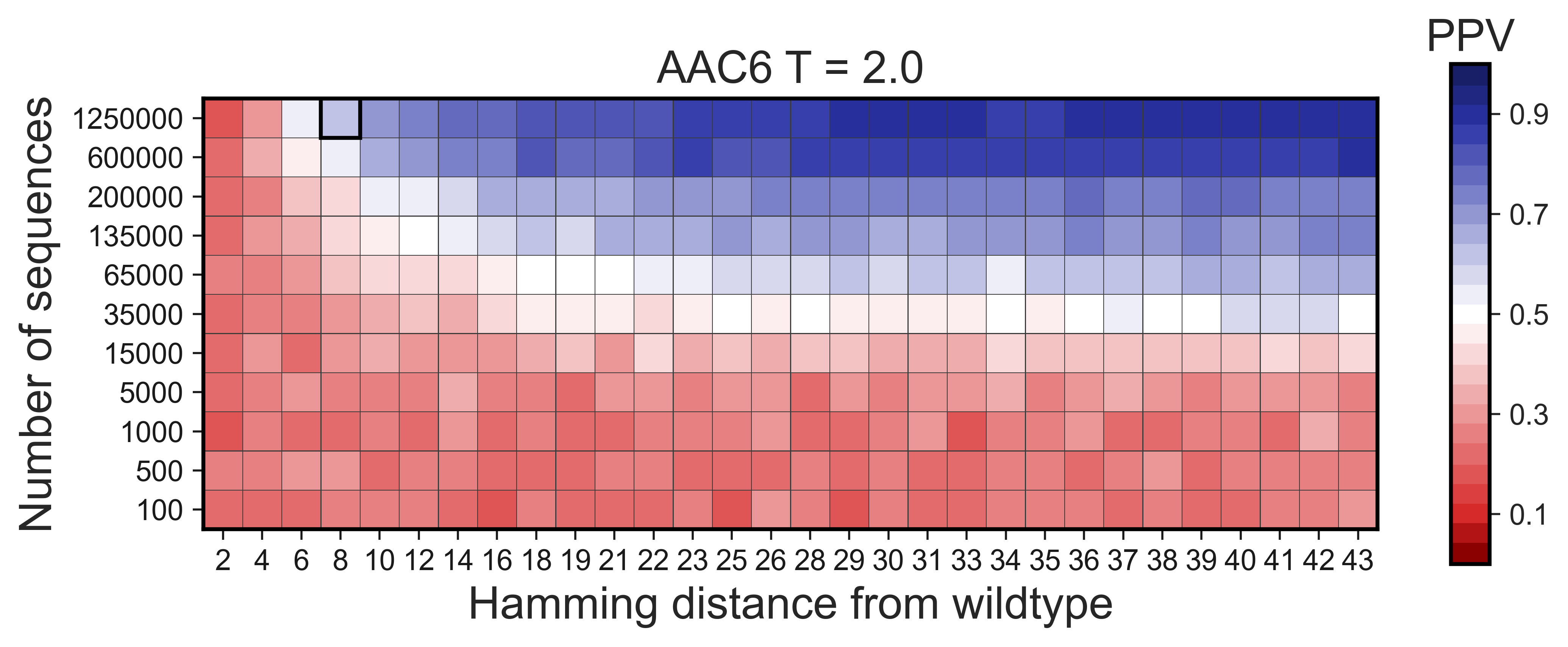}\vspace*{0.5cm}
\caption{\label{fig:Phase_diag_AAC_1_6}   {\bf Accuracy of contact prediction in dependence of sequence number and sequence divergence:} The figure shows the accuracy of contact prediction as a function of the average sequence divergence from wildtype AAC6 and the depth of the sequenced library, for selection strength $T=2$. The accuracy is measured via the positive predictive value (PPV), {\em i.e.}, the fraction of true positive contact predictions in the first 55 DCA-predicted contacts, cf.~{\em Methods} for details.  The highlighted square indicates an average Hamming distance of about 8 and a sequence library of 1,250,000, as realized in \cite{stiffler2020protein}.}
\end{center}
\end{figure*}

The AAC6 results show that reduced sequence divergence can, at least partly, be compensated by a strong increase in the number of sequences in the evolved MSA, cf.~Fig.~\ref{fig:Phase_diag_AAC_1_6}, which confirms original findings of \cite{stiffler2020protein}. Even if having only an average Hamming distance of about 8 substitutions, the large library of more than $10^6$ sequences allows for the detection of a weak contact-related signal.

\begin{figure*}[htb!]
\begin{center}
\includegraphics[width=0.8\textwidth]{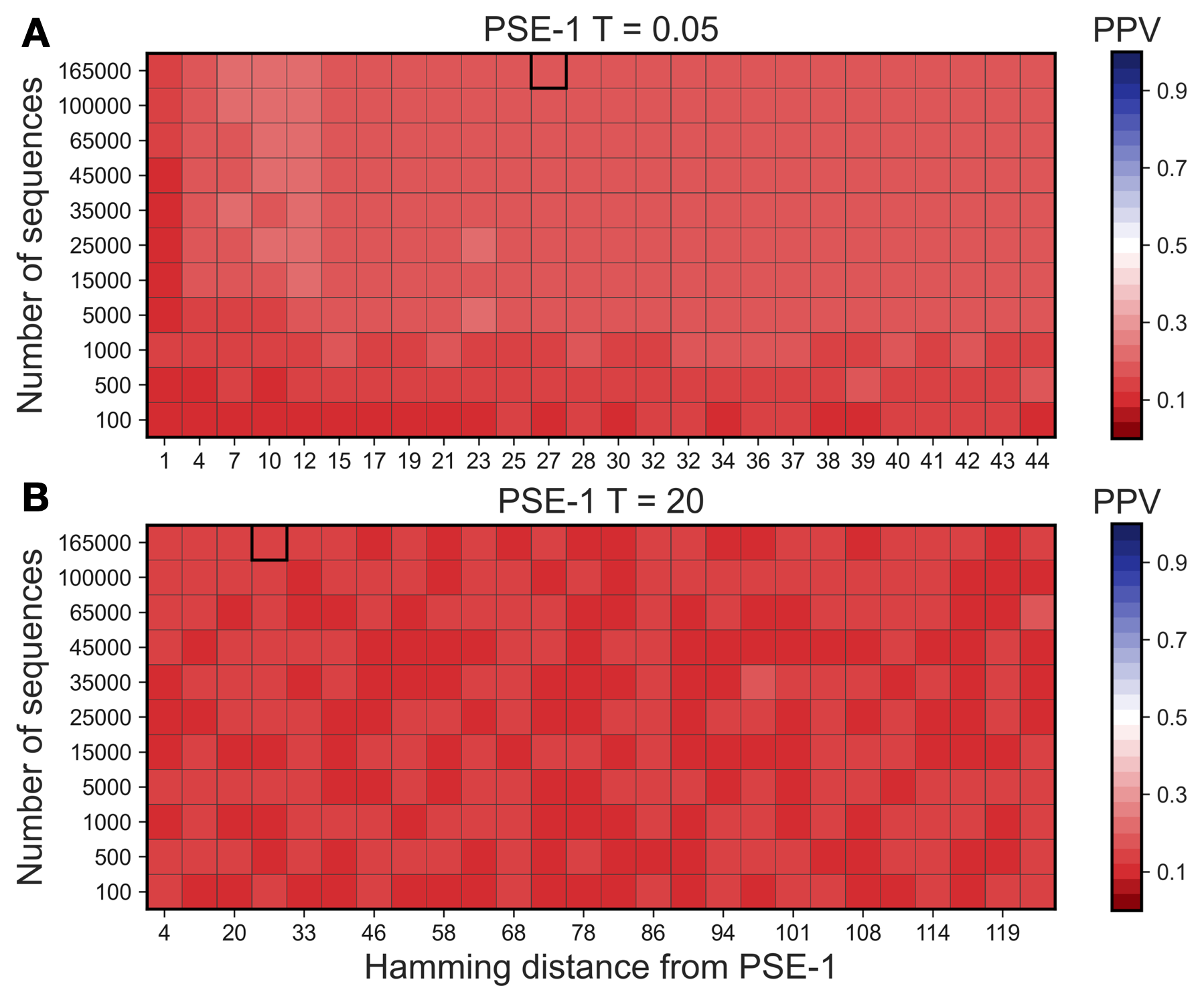}\vspace*{0.5cm}
\caption{\label{fig:phase_diag_PSE_lowhighT}   {\bf Accuracy of contact prediction in dependence of sequence number and sequence divergence:} The panels show, for the case of very strong selection ($T=0.05$, Panel A) and very weak selection ($T=20$, Panel B), the accuracy of contact prediction as a function of the average sequence divergence from wildtype PSE-1 and the depth of the sequenced library. The accuracy is measured via the positive predictive value (PPV), {\em i.e.}, the fraction of true positive contact predictions in the first $100$ DCA-predicted contacts, cf.~{\em Methods} for details.  The highlighted square indicates an average Hamming distance of about 27 and a sequence library of 165,000, as realized in \cite{stiffler2020protein}.}
\end{center}
\end{figure*}

The results depend substantially on the strength of selection. Fig.~\ref{fig:phase_diag_PSE_lowhighT} shows the extreme cases of very strong and very weak selection discussed before. Both show inaccurate prediction. An important difference becomes visible when looking at the horizontal axes: all use the same number of simulated evolutionary steps. In the case of strong selection, sequences stay closer to the wildtype, since most mutations are deleterious and selected against, and they stay close to each other. So while being all functional, they do not accumulate sufficient sequence variability to provide a reliable epistatic signal. In the case of extremely weak selection, almost all mutations are acceptable. Sequences are found to diverge strongly from the initial PSE-1 sequence, but the absence of selection causes also an absence of coevolution. 

\section{Discussion\label{sec:Discussion} }

The aim of this work was to showcase the potential of evolutionary models in data-driven sequence landscapes. Recent progress in landscape modeling has led to advances in using sequence alignment to predict protein structure, mutational effects and even to design non-natural but biologically functional sequences. Here we show that, equipped with a simple stochastic dynamics capturing the interplay between mutation and selection, these landscapes lead to models which are able to describe in a quantitatively accurate way the results of evolution experiments. This is not only restricted to proteins, as studied in this work, but similar evolution experiments have been performed for RNA \cite{zhou2018global} and could therefore be analyzed in an analogous way starting from sequence landscapes for RNA families \cite{kalvari2021rfam}.

\begin{figure*}[htb!]
\begin{center}
\includegraphics[width=0.8\textwidth]{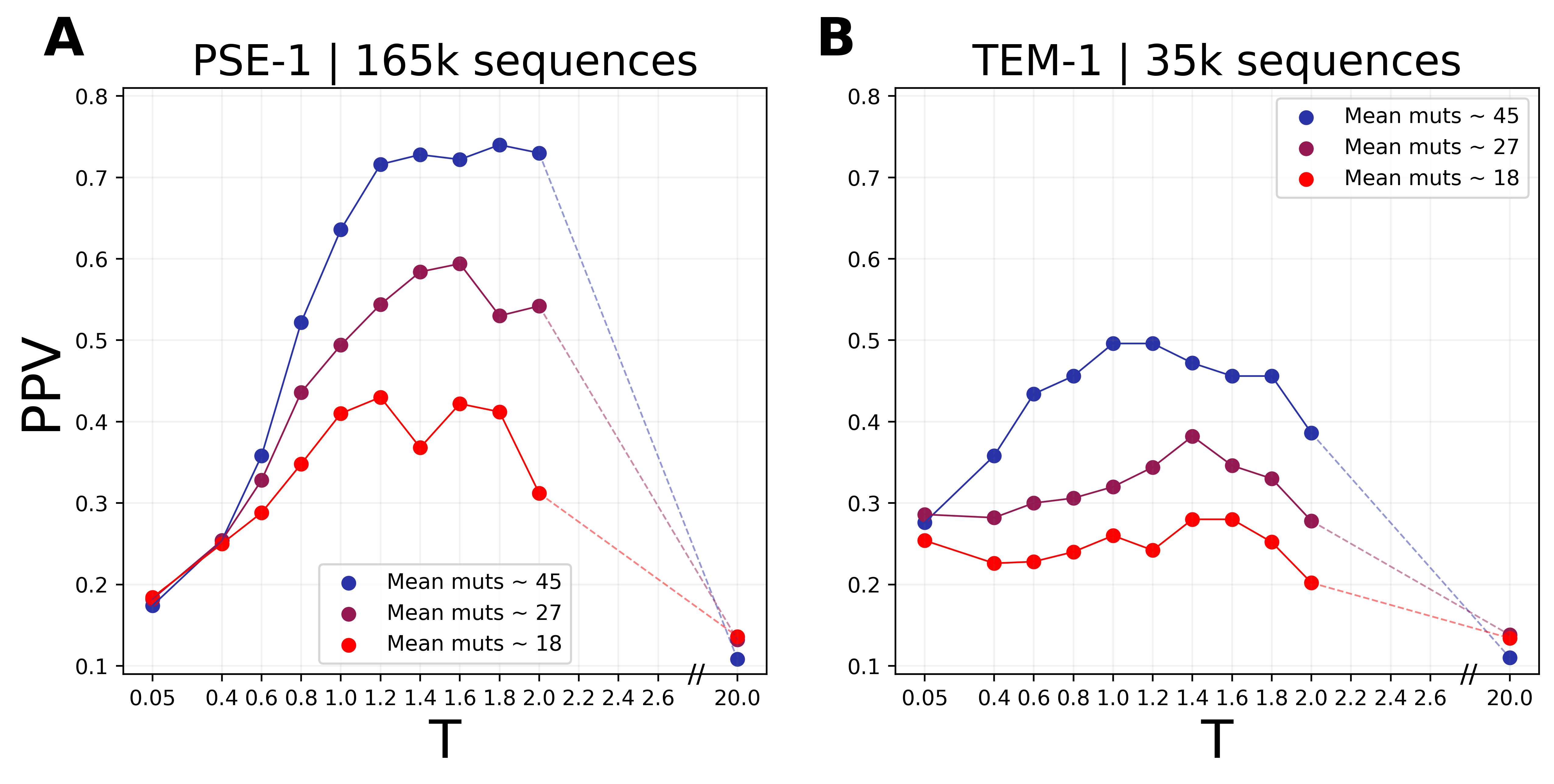}\vspace*{0.5cm}
\caption{\label{fig:PPV_T}   {\bf Dependence of the contact-prediction accuracy on selection strength:} We show the PPV (100 predicted contacts)  of simulated MSAs at variable selection strength $T$ (Panel A for PSE-1, Panel B for TEM-1), and for different sequence distances from the wildtype protein. We predict that, for the distances observed in the evolution experiments (27 for PSE-1, 18 for TEM-1), both experiments would have benefited from slightly lower anti-biotic concentrations. }
\end{center}
\end{figure*} 

The applications for experimental evolution are evident: we can use our modeling to optimize experimental evolution protocols, e.g., when we search for fully functional sequences but at some minimum number of mutations from a starting sequence, or when we want to explore sequence space optimally for contact prediction. In this case, we could, e.g., optimize the selection strength. In the case of the beta-lactamases studied in this article, Fig.~\ref{fig:PPV_T} shows that a slightly lower selection pressure ({\em i.e.} higher selection temperature) would have led to even better contact predictions. However, this potential increase is weak as compared to the one reachable by more diverged sequences.

A possible obstacle in such applications is the fact that the selection temperature $T$, which we use to model selective pressure, has to be fitted from experimental data via the slope of the statistical energies of the evolved sequences vs. their distance from wildtype. To understand the minimal sequence requirements for reaching robust and accurate slope estimates, we have subsampled the experimental sequence libraries of PSE-1 for rounds 10 and 20. As is shown in Supplementary Fig.~\ref{fig:subsample}, we observe (i) that the slope can be estimated accurately already from about 200-300 sequences, while the estimation error becomes large when using less than 100 sequences, and (ii) that the estimates are almost equal for round 10 and round 20. We conclude that the selection temperature $T$ can be reliably determined with moderate experimental effort (low number of sequences, few experimental rounds). Once estimated, the parameters can be used in simulations, which may guide more massive experiments evolving large sequence libraries over many rounds.

We see our current model as a starting point for more detailed evolutionary models. There is space for a substantial gain in accuracy: we can introduce biases in the mutations introduced by error-prone PCR directly into the model \cite{moore2000modeling,pritchard2005general}, the latter can be derived from data by analyzing synonymous mutations. Furthermore, we can introduce codon bias, the difference between transitions and transversions, the fact that error-prone PCR may introduce simultaneously several mutations before selection, or the emergence of phylogeny in cycles of mutation and selection. 

The modeling can also benefit from experimental feedback. If sequence libraries would also be sequenced before and after the selection step, we could establish a better correspondence between statistical energies and selection, up to a gauge of statistical energies vs. antibiotic concentrations.

However, the potential of such evolutionary models in data-driven landscapes goes far beyond the application to experimental evolution. As is shown by SEEC \cite{de2020epistatic}, already the simplest non-trivial evolutionary model allows for illuminating important consequences of epistasis in evolution, like the site- and time-dependence of substitution rates. We anticipate that the proposed modeling framework may capture many of these effects in a highly quantitative way. The relatively simple modeling framework proposed in our paper might also be a starting point for more theoretical-mathematical analyses about, e.g., the emergence of epistatic signals in sequence libraries. In this context, it might also be interesting to see in how far more distributed signatures of epistatic signal, possibly related to protein function rather than contacts, become visible in experimentally evolved sequence libraries, cf.~\cite{rivoire2016evolution,tubiana2019learning,shimagaki2019selection}.

\section{Methods\label{sec:Methods}}

\subsection{Sequence data\\}

\subsubsection{\label{sec_experimental}Sequences from experimental evolution}

We include in our analysis the sequence data coming from the experiments of \textit{in vitro} evolution by \cite{fantini2020protein} on TEM-1 and by \cite{stiffler2020protein} on PSE-1 and AAC6. 

The aligned amino-acid sequences from \cite{fantini2020protein} were kindly provided by the authors prior to publication, and can also be found at http://laboratoriobiologia.sns.it/supplementary-mbe-2019/. The raw sequencing reads are available at the National Centre for Biotechnology Information Sequence Read Archive (SRA) with accession code PRJNA528665 (http://www.ncbi.nlm.nih.gov/sra/PRJNA528665). Amino-acid sequences with more than $6$ gaps were discarded as a quality control to remove sequences with lower quality. 

\cite{stiffler2020protein} ran two experiments using the PSE-1 beta-lactamase and the AAC6 acetyltransferase as starting wildtypes.  Aligned sequencing reads from the last round of the two experiments (translated into amino-acid sequences) can be found at https://github.com/sanderlab/3Dseq.
The raw sequencing reads are available at the National Centre for Biotechnology Information Sequence Read Archive (SRA) with accession code PRJNA578762 (http://www.ncbi.nlm.nih.gov/sra/PRJNA578762).

Our models are built for the Pfam-annotated positions using the corresponding Pfam domains PF13354 (Beta-lactamase2) and PF00583 (Acetyltransf$\_$1). We re-aligned the wildtype sequence using the hmmalign command from the HMMer software suite \cite{eddy2011accelerated} and profile Hidden Markov Models (pHMM) downloaded from Pfam \cite{mistry2021pfam}. We then removed from the experimental MSA all columns corresponding to non-matched states of the wildtype sequence. 

The resulting MSAs of experimentally evolved sequences have 202 sites and 165,855 sequences for PSE-1 (round 20), and 34,431 sequences for TEM-1 (generation 12). For AAC6 we find 117 sites and 1,260,048 sequences (round 8). 

\subsubsection{Natural homologous sequences and preprocessing of the training set}

The MSAs of natural homologous sequences of the two considered protein families PF13354 (Beta-lactamase2) and PF00583 (Acetyltransf$\_$1) were generated running the hmmsearch command from the HMMer software suite \cite{eddy2011accelerated} on the UniProt database \cite{uniprot2021uniprot}. Insertions were removed, and sequences with more than $10\%$ gaps and duplicated sequences were excluded to improve the quality of the alignment. Any sequence closer than $80\%$ to the wildtypes TEM-1, PSE-1, or AAC6 was excluded from the alignments to avoid the introduction of biases towards these sequences in the bmDCA learning. The resulting MSAs included $18,333$ ($43,576$) homologous and non-identical aligned sequences of length $202$ ($117$) for PF13354 (PF00583).

Note that some residues, which are present in the N- and C-terminal regions of the experimental sequences, are not covered by the Pfam domains, and therefore excluded from our analyses. Extending the MSA beyond the borders of the Pfam domains would lead to the inclusion of evolutionarily less conserved positions, and thus to the inclusion of highly gapped columns into the MSA of natural data. Such columns have been previously found to compromise the accuracy of DCA landscapes \cite{figliuzzi2016coevolutionary} and are therefore left out in this study.

The natural MSA were used to train two Potts models using bmDCA \cite{figliuzzi2018pairwise} in the implementation of \cite{barrat2020sparse}, which provides the currently most accurate DCA models. 

\subsection{Evolutionary model}

\begin{figure*}[htb!]
\begin{center}
\includegraphics[width=\textwidth]{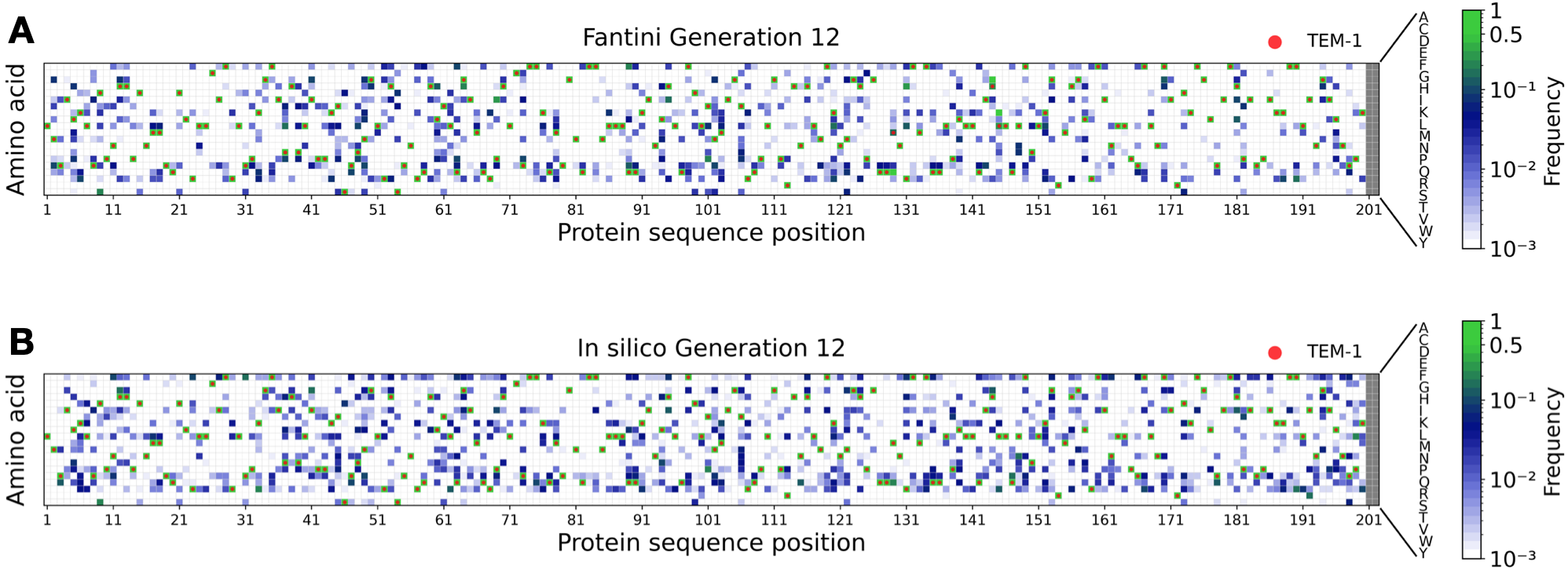}\vspace*{0.5cm}
\caption{\label{fig:mutation_spectrum_TEM1}   {\bf Position specific amino-acid frequencies for experimental and simulated sequence libraries:} Panel A shows the frequencies $f_i(a)$ of usage of amino acid $a$ in site $i$ in round 12 of experimental TEM-1 evolution, Panel B shows the same quantity for simulated evolution. Both plots have a Spearman rank correlation of 79\%.}
\end{center}
\end{figure*}

\begin{figure*}[htb!]
\begin{center}
\includegraphics[width=\textwidth]{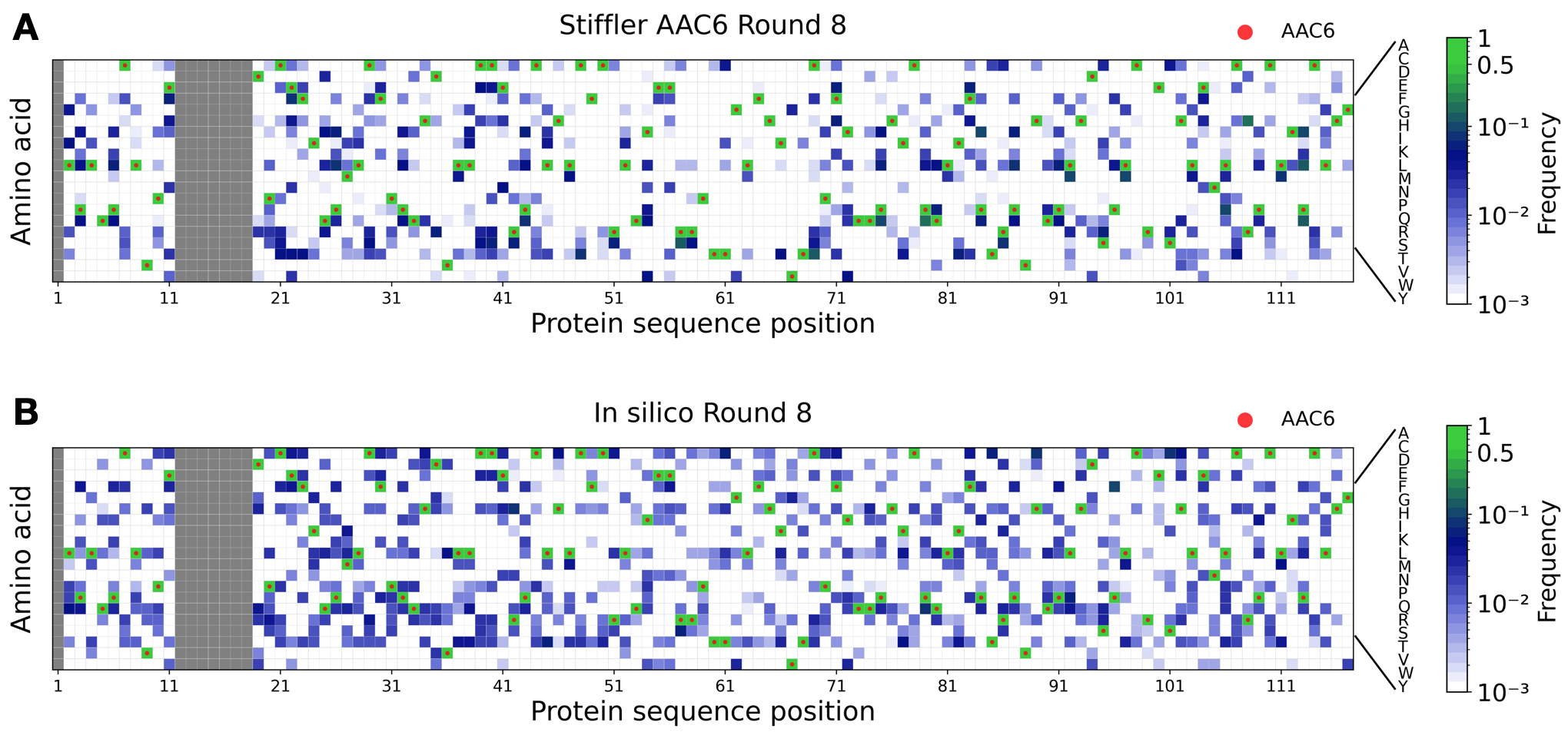}\vspace*{0.5cm}
\caption{\label{fig:mutation_spectrum_AAC}   {\bf Position specific amino-acid frequencies for experimental and simulated sequence libraries:} Panel A shows the frequencies $f_i(a)$ of usage of amino acid $a$ in site $i$ in the experimental AAC6 evolution after round 8, Panel B shows the same quantity for simulated evolution. Both plots have a Spearman rank correlation of 77\%.}
\end{center}
\end{figure*}

As already discussed in {\em Results}, our evolutionary model combines mutations at the nucleotide level with selection at the level of aligned amino-acid sequences. We therefore need to specify both the nucleotide sequence ${\bf n} = (n_{11},n_{12},n_{13},...,n_{i1},n_{i2},n_{i3},...,n_{L1},n_{L2},n_{L3})$ and the resulting amino-acid sequence ${\bf a} = (a_1,...,a_L)$, which is translated from ${\bf n}$ using the standard genetic code. Since we consider full-length aligned sequences of Pfam domains, stop codons are not allowed in ${\bf n}$. Furthermore, we have to accommodate alignment gaps possibly existing in ${\bf a}$: a gap in ${\bf a}$ is represented by a triplet of gaps in ${\bf n}$. Gaps are not changed during our simulations, our model does consider only single-nucleotide substitutions, but no insertions and no deletions. Note that the grey columns in Figs.~\ref{fig:mutation_spectrum},
\ref{fig:mutation_spectrum_TEM1}, \ref{fig:mutation_spectrum_AAC} correspond to gaps in the wildtype sequence, which are conserved both in the experiment and in the model.

As mentioned before, for each codon $(n_1,n_2,n_3)\in \{A,C,G,T\}^3$, we consider the set of amino acids ${\cal A}_{acc}(n_1,n_2,n_3) \subset \{A,...,Y\}$, which are accessible from $(n_1,n_2,n_3)$ by a single nucleotide mutation. 

Our simulation of sequence evolution proceeds by iterating the following three steps defining a Markov chain (MC) in the space of nucleotide sequences (note that, due to the degeneracy of the genetic code, the process is {\em not} a Markov chain in amino-acid sequence space):
\begin{enumerate}
\item A position $i \in \{1,...,L\}$ is chosen uniformly at random along the amino-acid sequence, corresponding to the codon ${\bf n}_i=(n_{i1},n_{i2},n_{i3})$ and the amino acid $a_i$. While $a_i="-"$, {\em i.e.} a gap is chosen, we repeat the selection of the position $i$.
\item Out of all accessible amino acids $b\in {\cal A}_{acc}({\bf n}_i)$, we selected one using the conditional probability $P_\beta(b | {\bf a}_{-i} )$, which couples the amino acid $b$ explicitly to the sequence context ${\bf a}_{-i} = (a_1,...,a_{i-1},a_{i+1},...,a_L)$ :
\begin{equation}
    \label{eq:p_cond}
    P_\beta(b | {\bf a}_{-i} ) = \frac {\exp\left\{ \beta h_i(b) + \beta \sum_{j (\neq i)} J_{ij}(b,a_j)
    \right\}}{z_i({\bf a}_{-i} )} \ ,
\end{equation}
with 
\begin{equation}
    \label{eq:z_cond}
z_i({\bf a}_{-i} ) = 
\!\!\!\!\!
\sum_{b\in {\cal A}_{acc}({\bf n}_i)} 
\!\!\!\!\!
\exp\Big\{ \beta h_i(b) + \beta \sum_{j (\neq i)} J_{ij}(b,a_j) 
\Big\}
\end{equation}
being a normalization constant. In difference to $Z$ in Eq.~\eqref{eq:P}, it can be calculated efficiently by summing over the less than 20 accessible amino acids.
\item One out of the possible codons for amino acid $b$, which differs from ${\bf n}_i$ in a single nucleotide, is selected uniformly at random.
\end{enumerate}
The new amino acid $b$ substitutes $a_i$ in ${\bf a}$, and the new codon ${\bf n}_i$ in ${\bf n}$. We thereby conserve the coherence between nucleotide and amino-acid sequence.

To simulate an entire MSA of $M$ sequences, the process is initiated $M$ times in the wildtype reference sequence, and $M$ independent runs of the MC are performed. The number of steps in these MCs is chosen such that the average Hamming distance of the generated amino-acid sequences reaches a target number. Note that the Hamming distances may vary from MC to MC, since ${\cal A}_{acc}({\bf n}_i)$ contains the case $b=a_i$ accessible via any synonymous mutations. The Hamming distance can therefore assume any value between zero and the number of performed mutational steps.

\subsection{Simulated sequence data for contact prediction}

Our evolutionary algorithm has three input parameters adding to the wildtype sequence and the statistical-energy model: the number of sequences $M$, the number $N_{MC}$ of steps of our evolutionary Markov chain model, and the selection temperature $T$. Given this triplet of numbers it outputs an MSA obtained simulating evolution for $N_{MC}$ iterations starting from the wildtype sequence, repeating the sampling independently $M$ times at temperature $T=1/\beta$.

For each wildtype sequence, we simulated the outcome of different protein evolution experiments by scanning these three input parameters within a range of interest. For MSA generated starting from TEM-1 or PSE-1 (AAC6) we varied $M$ in the range $100-165,000$ ($500-1,250,000$), $N_{MC}$ in the range $5$-$255$ ($4$-$120$) and $T$ in the range $0.05$-$20$. 

To save resources and time, given the computational cost of sampling, we opted for a scheme that would allow us to reduce the number of independent MC chains needed to simulate evolution. For each temperature $T$, we run $165000$ ($1250000$) independent Markov Chains for TEM-1 and PSE-1 (AAC6) and printed MSAs at the desired number of MC steps until $255$ ($120$) MC steps. The MSAs with less sequences were obtained by randomly subsampling without replacement from the MSA with $165000$ ($1250000$) sequences. To produce more statistics we ran the same simulations $5$ times.

\subsection{Contact prediction}

Contact prediction was performed using GaussDCA \cite{baldassi2014fast} for all MSA, included, for coherence, the experimental ones. GaussDCA is the computationally most efficient implementation of DCA. Its accuracy of contact prediction is slightly inferior to plmDCA or bmDCA. However, we use it since in our analysis we had to predict contacts for a large number of partially deep simulated MSA (cf.~Fig.~\ref{fig:phase_diag}) to explore multiple combinations of sampling time, sample size and selection strengths. 

The reweighting parameter was set to $0$ for contact prediction of {\em in-silico} MSAs, as this reduces computational time and is coherent with the independence of the simulated Markov chains. On the other hand, contact prediction of experimental MSAs was performed using the default option ":auto" of GaussDCA for reweighting. These different treatments of simulated and experimental sequences are based on the fact, that simulations generate statistically independent sequences (conditioned to wildtype initialization), while the experiments may generate sequence ensembles having non-trivial phylogenetic effects. The pseudocount was set to $0.6$ ($0.5$) for PSE-1 and TEM-1 (AAC6) empirically, as we found it to be a good intermediate value for MSAs with very different statistics.

Intra-chain atomic distances for both families were obtained by running the single-protein mode of the code provided by Pfam Interactions (https://doi.org/10.5281/zenodo.4080947), we used the shortest distance between heavy atoms of the two amino acids among all structures of the Protein Data Bank (PDB) \cite{burley2021rcsb} listed in Pfam. Following standards in coevolutionary contact prediction, all pairs with distance below 8\AA \,and a minimum separation of 5 positions along the sequence are kept as contacts for the calculation of the PPV (positive predictive value). For AAC6 we used a more stringent cutoff of 5.5\AA, since the structural variability across the protein family is already well represented in the PDB.

\section*{Acknowledgments}
We are grateful to the authors of \cite{fantini2020protein}, who shared their data with us prior to publication, and in particular M. Fantini, A. Pastore and P. de los Rios for interesting discussions about our first results. We also thank Anna Paola Muntoni for contributing to the implementation of bmDCA used in this work, and  to Giancarlo Croce, Kai Shimagaki, Jeanne Trinquier, Simona Cocco and Remi Monasson for useful discussions.
This work was partially funded by the EU H2020 Research and Innovation Programme MSCA-RISE-2016 under Grant Agreement No. 734439 InferNet (Martin Weigt), and by a grant from the Simons Foundation (\#454955, Francesco Zamponi).

\section*{Data and code availability}
The code for evolutionary simulations is available on Github (https://github.com/matteobisardi/SeqEvol) in the form of a Julia package. Data necessary to reproduce the results can be found in a Github repo (https://github.com/matteobisardi/DataSeqEvol), this includes wildtype sequences, bmDCA parameters and in-silico generated MSAs.

\bibliographystyle{natbib}
\bibliography{refs}


\onecolumngrid
\beginsupplement

\section*{Supporting information}

\begin{figure*}[htb!]
\begin{center}
\includegraphics[width=0.8\textwidth]{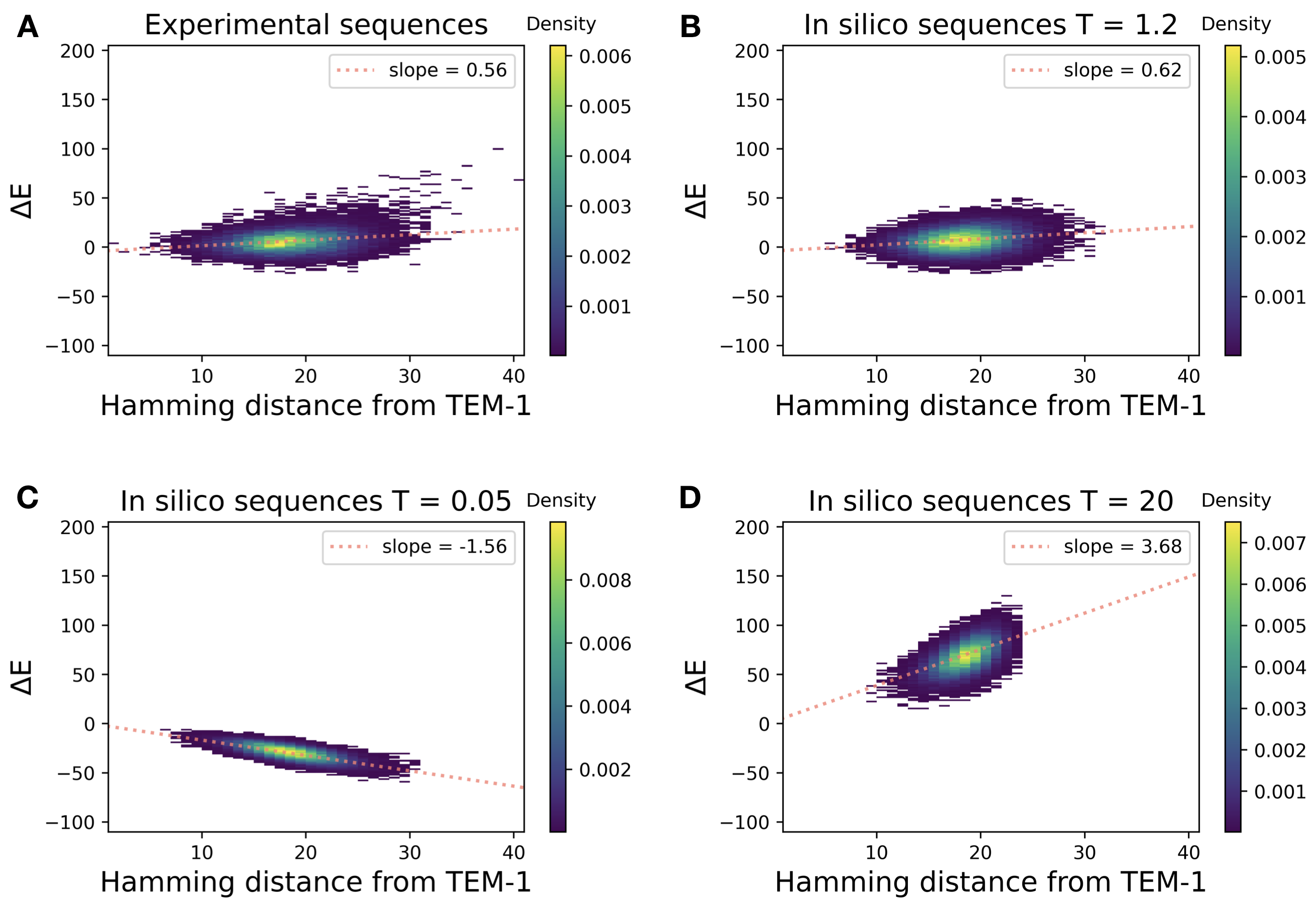}\vspace*{0.5cm}
\caption{\label{fig:E_of_d_tem1}   {\bf Statistical energy in dependence of sequence distance from wildtype TEM-1:} Panel A shows the statistical energies of the sequences from generation 12 in Fantini et al., in dependence of the Hamming distance (number of substituted amino acids) from the wildtype TEM-1. Panel B shows the same quantities for the simulated sequences, where selection strength $T$ and the number of simulated evolutionary steps are adjusted to reproduce the average distance and the slope from Panel A. Panel C shows an example of strong selection ($T\ll 1$) leading to optimized sequences having lower statistical energies / higher fitness. Panel D shows the case of very weak selection ($T\gg 1$) resulting in random, mostly deleterious substitutions strongly increasing statistical energy.}
\end{center}
\end{figure*} 

\begin{figure*}[htb!]
\begin{center}
\includegraphics[width=0.8\textwidth]{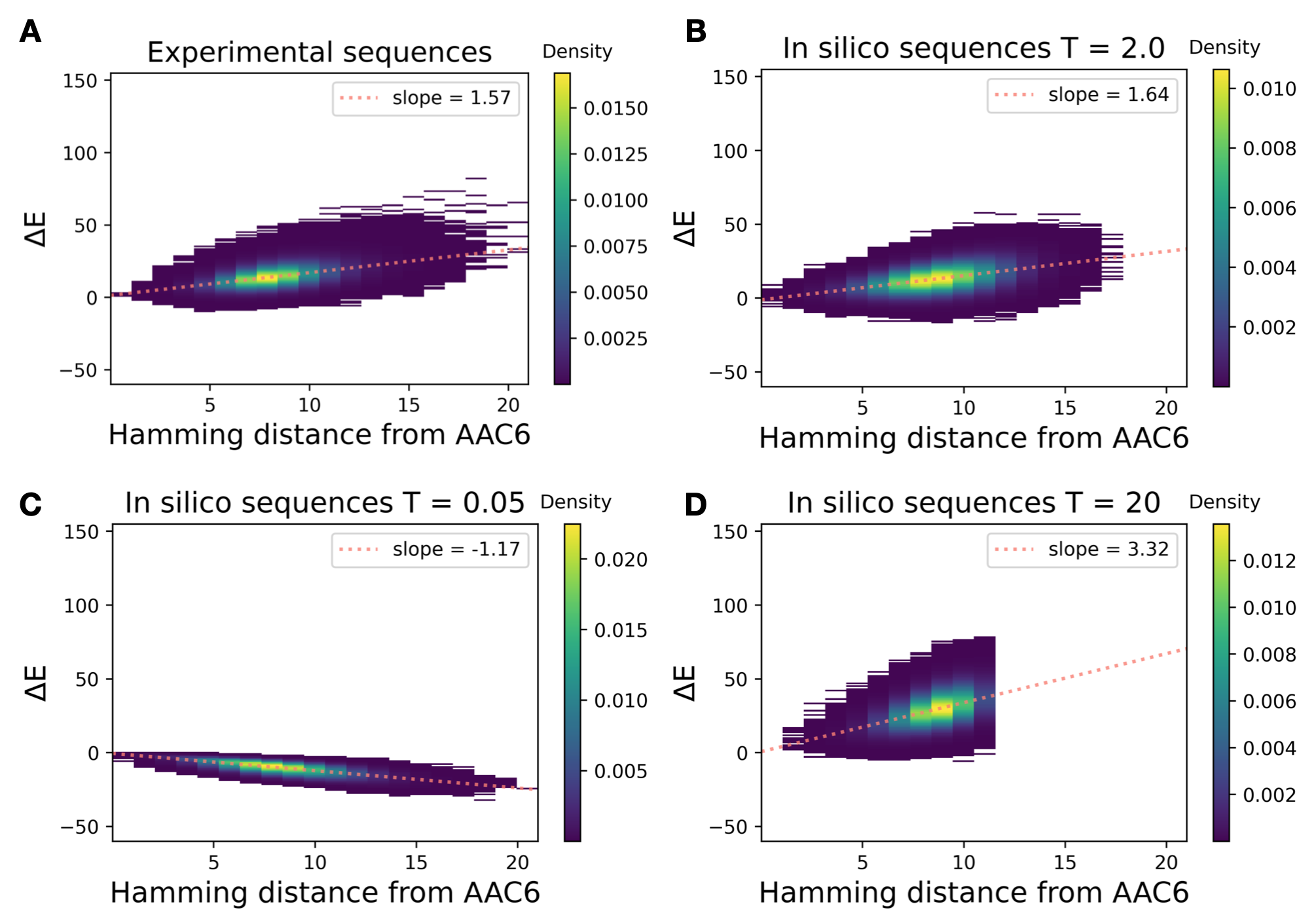}\vspace*{0.5cm}
\caption{\label{fig:E_of_d_aac}   {\bf Statistical energy in dependence of sequence distance from wildtype AAC6:} Panel A shows the statistical energies of the sequences from round 8 in Stiffler et al., in dependence of the Hamming distance (number of substituted amino acids) from the wildtype AAC6. Panel B shows the same quantities for the simulated sequences, where selection strength $T$ and the number of simulated evolutionary steps are adjusted to reproduce the average distance and the slope from Panel A. Panel C shows an example of strong selection ($T\ll 1$) leading to optimized sequences having lower statistical energies / higher fitness. Panel D shows the case of very weak selection ($T\gg 1$) resulting in random, mostly deleterious substitutions strongly increasing statistical energy.}
\end{center}
\end{figure*} 

\begin{figure*}[htb!]
\begin{center}
\includegraphics[width=0.5\textwidth]{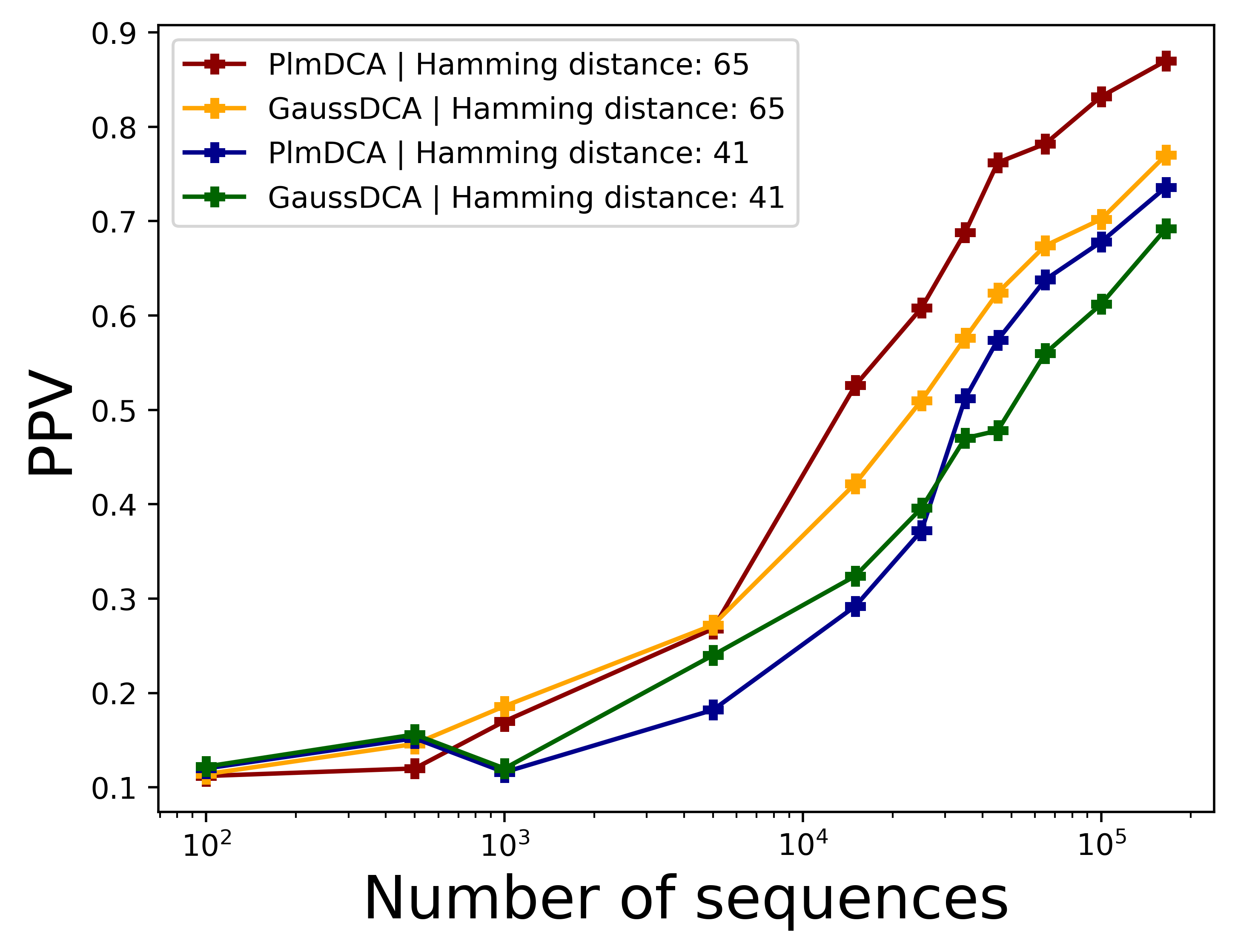}\vspace*{0.5cm}
\caption{\label{fig:phase_diag_plm}   {\bf Accuracy of contact prediction in dependence of sequence number:} The figure compares the accuracy of contact prediction of plmDCA vs. GaussDCA as a function of the sequence number, for two distances from wildtype PSE-1. The accuracy is measured via the positive predictive value (PPV), {\em i.e.}, the fraction of true positive contact predictions in the first $100$ DCA-predicted contacts, cf.~{\em Methods} for details. The selection strength $T=1.4$ corresponds to the experimental condition in~\cite{stiffler2020protein}.}
\end{center}
\end{figure*} 

\begin{figure*}[htb!]
\begin{center}
\includegraphics[width=0.8\textwidth]{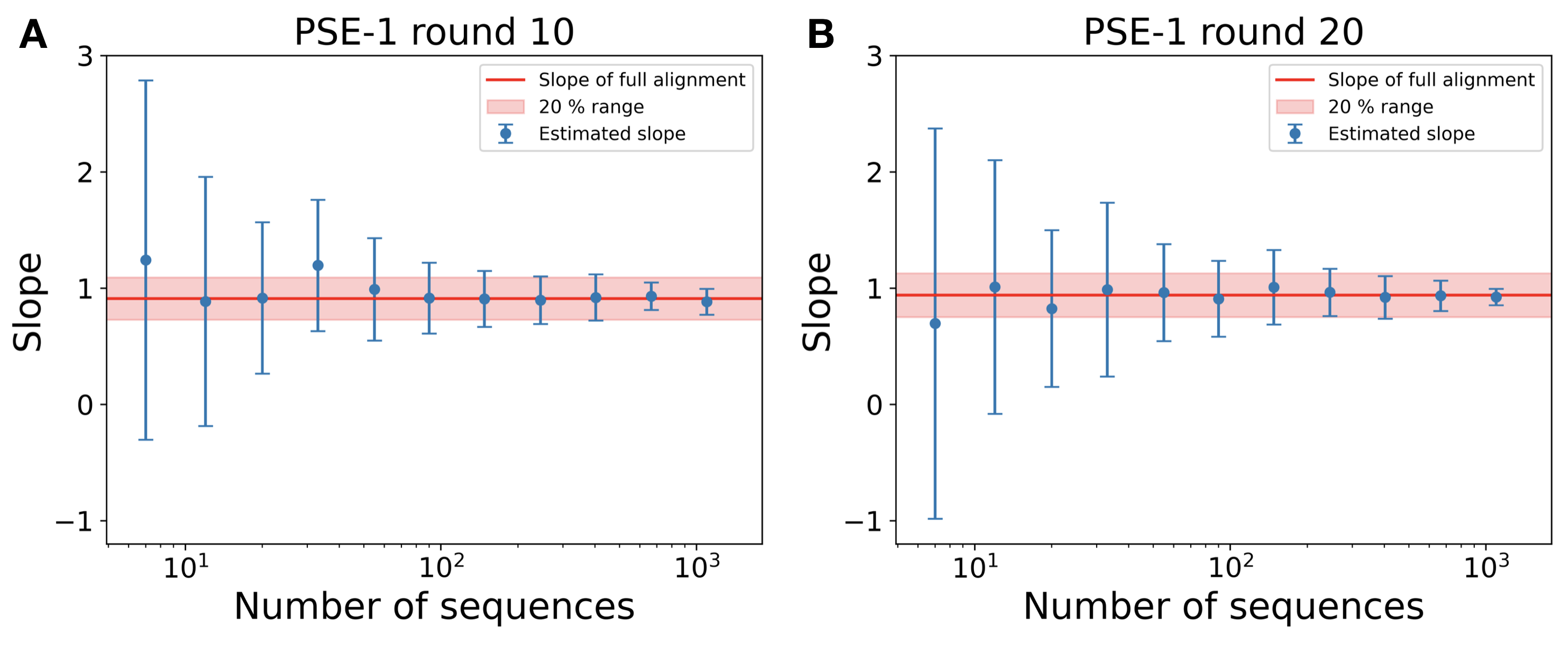}\vspace*{0.5cm}
\caption{\label{fig:subsample}  {\bf Slope of the statistical energy vs. sequence distance from wildtype estimated from subsamples of the PSE-1 sequence libraries:} The panels show the means and standard deviations of the estimated slopes obtained from subsamples of the experimental PSE-1 sequence libraries at round 10 (Panel A) and round 20 (Panel B). The values obtained for the full libraries are evidenced by the red horizontal line, together with a 20\%-interval. We observe that estimates fall reliably into this interval when at least 200-300 sequences are used, and that the estimated slopes are almost identical for the libraries obtained after 10 or 20 rounds of experimental evolution in \cite{stiffler2020protein}.}
\end{center}
\end{figure*} 

\end{document}